\PassOptionsToPackage{authoryear}{natbib}
\documentclass[linenumbers,twocolumn]{openjournal}

\newcommand{\uat}[2]{#1}

\newcommand{\orcidlink}[1]{\href{https://orcid.org/#1}{\includegraphics[width=8pt]{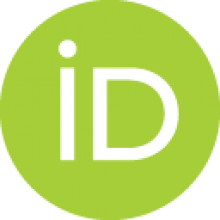}}}

\usepackage{booktabs, rotating, makecell, multirow, xspace,color}
\usepackage{hyperref}
\hypersetup{colorlinks,allcolors=black}
\newcolumntype{L}[1]{>{\raggedright\let\newline\\\arraybackslash\hspace{0pt}}m{#1}}
\newcolumntype{C}[1]{>{\centering\let\newline\\\arraybackslash\hspace{0pt}}m{#1}}
\newcolumntype{R}[1]{>{\raggedleft\let\newline\\\arraybackslash\hspace{0pt}}m{#1}}
\shorttitle{Dancing Streams In Merging Halos}
\shortauthors{Weerasooriya et al.}
\graphicspath{{./}{figures/}}

\usepackage{comment}

\usepackage{graphicx}
\usepackage{enumitem}
\setlist[enumerate]{align=left}

\begin{document}
\title{Dancing Streams In Merging Halos: Stellar Streams in a MW--LMC-like merger}

\author{Sachi Weerasooriya\,\orcidlink{0000-0001-9485-6536}$^*$}
\altaffiliation{$^*$sachiwee@gmail.com}

\affiliation{Carnegie Science Observatories, 813 Santa Barbara Street, Pasadena, CA 91101, USA}
\author{Tjitske Starkenburg\,\orcidlink{0000-0003-2539-8206}}
\affiliation{Department of Physics and Astronomy, Northwestern University, 2145 Sheridan Rd, Evanston IL 60208, USA}
\affiliation{Center for Interdisciplinary Exploration and Research in Astrophysics (CIERA), Northwestern University, 1800 Sherman Ave, Evanston IL 60201, USA}
\affiliation{NSF-Simons AI Institute for the Sky (SkAI), 172 E. Chestnut St., Chicago, IL 60611, USA}

\author{Emily C. Cunningham\,\orcidlink{0000-0002-6993-0826}$^{\dagger}$}
\altaffiliation{$^{\dagger}$NASA Hubble Fellow}

\affiliation{Department of Astronomy, Columbia University, 550 West 120th Street, New York, NY, 10027, USA }

\affiliation{Department of Astronomy, Boston University, 725 Commonwealth Ave., Boston, MA 02215, USA}

\author{Kathryn V Johnston\,\orcidlink{0000-0001-6244-6727}}
\affiliation{Center for Computational Astrophysics, Flatiron Institute, Simons Foundation, 162 Fifth Avenue, New York, NY 10010, USA}
\affiliation{Department of Astronomy, Columbia University, 550 West 120th Street, New York, NY, 10027, USA}

\begin{abstract}

Stellar streams --- formed from tidally stripped globular clusters or dwarf galaxies --- are sensitive tracers of a galaxy's accretion history and gravitational potential. While numerous streams are known in the Milky Way (MW) the formation and evolution of stellar streams has been primarily studied in isolated settings. The impact of subsequent galaxy interactions on stellar streams remains largely unexplored. Understanding merger-induced effects is however crucial given the accretion of the Large Magellanic Cloud (LMC) onto the MW, and the fact that for example M31 and Cen A have experienced recent mergers. We analyze the detailed evolution of 1024 stellar streams during a complete MW--LMC-like merger, systematically varying initial stream properties and considering various orbits for the infalling perturber. We find that an MW--LMC mass-ratio merger significantly alters stellar stream properties, including energy, angular momentum, orbit, and morphology. Some streams exhibit dramatic morphological changes or develop complex substructures, while others see substantial shifts in energy and/or angular momentum, or re-orient their orbital plane. Interestingly, strong morphological alterations do not necessarily correlate with large changes in energy or orbit. A few streams split apart with parts moving to different orbits, appearing disconnected in position and kinematics despite their common origin. Strong effects correlate with close encounters between stream particles and the infalling perturber at various times during the merger. Our findings highlight the considerable impact of significant accretion events on the properties of stellar streams, and the challenge to recover the initial orbits of streams from their appearance at the present-day. Visualizations of the detailed evolution of all 1024 stellar streams are available at \url{https://dancingstreamsinmerginghalos.github.io}.
\end{abstract}


\section{Introduction}\label{sec:intro}

Stellar streams are remnants of smaller stellar systems such as dwarf galaxies or globular clusters that have been torn apart by tidal forces \citep[e.g.][]{LBLB1995, Johnston+1995, Johnston+1996, Helmi1999, Johnston2016, Newberg2016, Helmi+2008, Helmi+2020}. As thin, elongated structures of previously accreted systems in which stars follow similar orbits, stellar streams are considered to be tracers of a galaxy's formation history as well as excellent probes of the local gravitational potential \citep{Johston1999, Helmi1999, Ibata+2001,Helmi2004,Johnston+2005,Fellhauer+2006,Law&Majewski2010,Bowden+2015,Bonaca&Hogg2018,Reino+2021,Pearson+2022, Ibata2024}. Streams have been used to estimate the mass distribution in the halo of the Milky Way \citep[MW,][]{Johston1999, Bonaca+2014,Gibbons+2014, Sanders2013, Koposov+2010,Kupper+2015, Reino2021, Dodd2022, Koposov2023, Ibata2024} and other galaxies \citep{Ibata+2004,Fardal+2013,Amorisco+2015}. Many stellar streams have been discovered in the MW during the last decades, a significant fraction of which are thin stellar streams that likely formed from globular clusters (e.g. \citealp{Odenkirchen+2001,Grillmair+2006,Koppelman2018, Ibata+2021,Malhan+2018,Ibata2024, Tian+2024}, see \citealp{BPW2024} for a review). For example, data from the Gaia mission 
has significantly increased the number of known stellar streams \citep{Koppelman2018, Gaia2016a,Gaia2016b,Gaia2018,Gaia2021a,Gaia2021b,Gaia2023, BPW2024}. Additionally, proper motions of stream stars from Gaia, particularly when combined with spectroscopic data, add opportunities for discovering new streams and constrain stream orbits, providing potential insights into their origin \citep[][among others]{Malhan2018, Fardal2019, Shipp+2019, Li2022, Martin2022, Viswanathan2023, BPW2024}.

The thinnest streams, formed from globular clusters, are the most sensitive to local changes in the gravitational potential. For example, several studies have shown that physical underdensities or gaps in thin streams may have formed from impacts by dark matter subhalos \citep[e.g.][]{Johnston2002, Ibata2002, Yoon2011, Carlberg2012, Ngan&Carlberg2014,Erkel+2016,Banik+2018,Carlberg&Grillmair2013,Carlberg2016,Ibata+2016,Erkel+2017,Bovi+2017,deBoer+2018,Bonaca+2020}. Off--stream stars, spurs, or bifurcations can also be indicative of subhalo interactions. \citep{Bonaca+2019} shows that a subhalo interaction may explain the presence of substructure in GD-1 stellar stream, but there may be valid alternative explanations as well \citep{DeBoer2020, Bonaca2020, Ibata2020, Dillamore+2022, Qian+22, Valluri2024}. \cite{Banik+2019} argue that Pal 5, a stream close to the center of the MW, is a poor constraint of the dark matter structure, and that streams in the outer halo will be needed to detect dark matter subhalos. Interestingly, recent results have highlighted many more intriguing substructures in stellar streams \citep{Ferguson2022}, and even stellar streams that appear to split into multiple components \citep{Li2021, Bonaca2019, Nidever2023}. These results reflect that the morphological and dynamical evolution of stellar streams may be quite complex.

\begin{figure*}[t]
        \includegraphics[width=\textwidth]{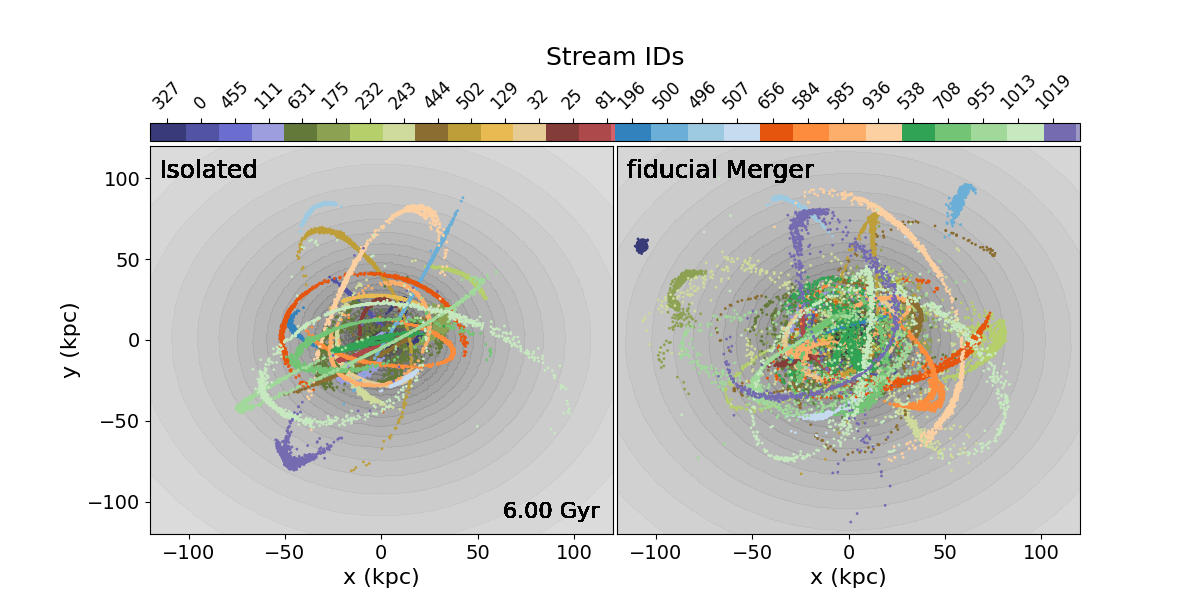}
        
       \caption{Positions for star particles from 27 of our full set of stellar streams evolved for $6\,\mathrm{Gyr}$ in a MW-like halo, in isolation (left) and during our 1:5 fiducial merger (right). The potential of the halos is shown in grey contours, and streams are color-coded by their stream ID to highlight individual streams. The following figures use a different color scheme where streams are grouped by orbital properties. A complete movie of this figure is available at \url{https://dancingstreamsinmerginghalos.github.io}.}
    \label{fig:movie}
\end{figure*} 

In particular, galaxies formed in a cosmological context are known to undergo mergers throughout their history, yet most studies of thin stellar stream evolution have only looked at isolated parent galaxies or the infall stage of accreted satellites, such as the LMC.
Indeed, the MW stellar halo is thought to be largely formed by the merging of the proto--MW with a relatively massive satellite, which may have contributed up to 20\% of the MW's present day dark matter and 50\% of its stellar halo \citep[e.g.][]{Ibata+1994,Helmi+1999,Chiba+2000,Majewski+2003,Bell+2008,Newburg+2009,Nissen+2010,Belokurov+2018,Helmi+2018,Myeong+2019,Matsuno+2019,Koppelman+2019,Yuan+2020,Naidu+2020}. More recently, the MW experienced two relatively significant accretion events which have contributed negligibly to the smooth stellar halo: the Sagittarius dwarf galaxy \citep[Sgr; e.g.][]{Ibata+1995,Ibata+2001,Newberg+2002,Majewski+2003,Belokurov+2006,Kruijssen+2020}, the most massive tidal stream of our galaxy at the present-day \citep{Belokurov+2006,Belokurov+2014,Hernitschek+2017,Sesar+2017},  and the Large and Small Magellanic Clouds (LMC and SMC). Both of these interactions may affect stellar streams in the MW halo \citep[e.g.][]{Erkel+2019,Shipp+2021,Dillamore+2022, Koposov2023, Lilleengen2023, Woudenberg2023, Valluri2024}. 

Simulations find that while globular cluster stellar stream orbits trace the current potential, they can be affected by the evolution of their host potential \citep{Zhao1999, Penarrubia2006, Erkel+2019, Arora+23}. For example, MW streams formed more than 3 Gyrs ago can be altered by the Sagittarius dwarf galaxy depending on their initial orbits \citep{Dillamore+2022}, and action space clustering for dwarf galaxy stellar streams can change when experiencing a sufficiently large merger \citep{Arora2022}. The predicted number of streams affected by the MW--LMC merger depends on the total mass of the LMC and its distance \citep{Erkel+2019}. But measured perturbations of the Orphan stellar stream have provided constraints on the mass of the LMC \citep{Erkel+2019, Koposov2023, Lilleengen2023,Brooks+2024}. Modeling a large number of streams in a time-evolving potential of the MW--LMC interaction until the present day,  \citet{Brooks+2024b} find that outer halo streams exhibit changes in width and proper motion misalignment due to the infall of the LMC, and that streams that experience close encounters with the LMC will not correctly recover the spherical mass profile of the MW through action clustering methods \citep{Brooks+2024}. However, the length, asymmetry, and velocity dispersion of the streams show no significant differences.
In addition to direct effects of the merger, stream--subhalo interactions are predicted to increase $20-40\,\%$ in the presence of an LMC-like satellite galaxy due to interactions with infalling LMC satellites \citep{Arora+23}. Another indirect effect of a major merger on streams can be through changes in the orientation and angular momentum of the stellar disk, or disk tilting, which can produce both narrower or more diffuse streams \citep{Nibaur+2024}. \citet{Carlberg+2020} and \citet{Qian+22} both investigate the properties of stellar streams that are accreted from a satellite galaxy and show that this is an alternative explanation for some of the off-track features seen in stellar streams. 

All these studies show that the morphology, kinematics, and orbits of stellar streams can be affected by merger-induced time-dependent changes in their host potential.  This highlights the need to explore the evolution of stellar streams during and after galaxy interactions in more detail, for a statistical set of stellar streams where initial properties are varied systematically. Studying the effects of galaxy interactions on streams systematically could inform interpretations of stellar stream observations in the MW and, with upcoming facilities, in nearby galaxies such as M31 or Cen A, and whether these observed streams could statistically or individually provide evidence of the host galaxy's recent merger history.

Larger samples of stellar streams in the MW and other nearby galaxies will allow for statistical constraints on dark matter halo structure and dark matter subhalos. These samples may be reachable with upcoming facilities \citep{Sanderson+2019, Pearson+2019, Pearson+2022, Shipp+2023, Aganze2024}.
The Legacy Survey of Space and Time \citep[LSST,][]{LSST2009,Ivezic+2019,Akeson+2019} of The Vera C. Rubin Observatory (Rubin) will detect more low surface brightness streams in the MW halo \citep{Shipp+2023}. The Nancy Grace Roman Space Telescope \citep[Roman,][]{Spergel+2015} will have a field of view 200 times better than Hubble, higher angular resolution, and deep imaging sensitivity that would discover low surface brightness globular cluster stellar streams in external galaxies \citep{Pearson+2019,Pearson+2022, Aganze2024}. The Euclid Space Telescope's \citep[Euclid,][]{EUCLID2022,Smirnov+2023,Kelvin+2023} surface brightness limits will also increase the detectable extra galactic stellar streams by orders of magnitude \citep{Pearson+2019}. With this huge increase in observations, both in the amount of information available for individual stellar streams and in the number of stellar streams predicted to be discovered in the coming years, a
better understanding of the evolution of stellar streams
in diverse formation environments is imperative. 

In this study, we provide an overview of how a population of stellar streams evolves during a major perturbation such as a galaxy merger. We study a large set of stellar streams with widely varying orbital parameters, and compare their evolution in merging halos to that in an isolated dark matter halo. In particular, we highlight the changes in stellar streams and their properties after the perturber has completely merged with the host halo. Our simulation suite allows us to quantify stellar stream properties both statistically and individually. To emphasize the need for systematic and statistical studies, Figure~\ref{fig:movie} presents a visualization of 27 stellar streams from the simulation presented in this paper. A comparison of left and right panels reveals significant morphological differences between streams evolved in isolation vs. in a MW--LMC-like merger. The wide range of large and small effects in morphology visible in the evolving stellar streams motivates our in-depth study to characterize this evolution and its dependencies on the initial properties of the streams.

In Section~\ref{sec:methods}, we describe the details of our simulation setup. In Section~\ref{sec:results}, we present our results, and we discuss a set of noteworthy individual streams in Section~\ref{sec:individual}. Lastly, we discuss and summarize our results in Section~\ref{sec:discussion} and Section~\ref{sec:conclusions}, respectively. In future work we will describe the evolution of stellar streams in interacting potentials for varying merger mass ratios (Weerasooriya et al. in prep.).

\section{Methods}\label{sec:methods}

Our methods can be separated into two parts.
First, we generate stellar streams in a static halo potential to create a large sample of stellar streams with consistently varying properties (Section~\ref{method:static}). Subsequently, these streams are placed in live halos and evolved forward in N-body simulations, either in an isolated host dark matter halo or with the host halo experiencing a merger (Section~\ref{method:merger}). 
\subsection{Generating stellar streams in a static potential}\label{method:static}
 We build a set of 1024 globular cluster stellar streams  using the distribution function for generating mock stellar streams from \citet{Fardal+2015} as implemented in the \texttt{Gala} python package \citep{gala}. We set the time steps and the frequency of generating new star particles in the stream such that each final stellar stream exists of 1000 star particles irrespective of stream orbit or age. 
 
Each stream is evolved around a dark matter halo with a total mass of $1.57\times 10^{12}\,M_{\odot}$, a Hernquist potential \cite{Hernquist1990}, and a scale radius of $a = 40.85$~kpc based on \cite{GaravitoCamargo+2019}. We systematically vary the properties of each stream, including their total mass, apocenter radius, orbit circularity and age as described below. A summary of the initial conditions for the 1024 streams is presented in Table~\ref{table:tab1}.
\begin{enumerate}
\item[{\it Mass:}] We set the initial globular cluster total mass to either $M_{\star {\textrm{,tot}}} = 10^4\;M_{\odot}$ (512 streams) or $M_{\star {\textrm{,tot}}} = 10^6\;M_{\odot}$ (512 streams).

\item[{\it Apocenter Radius:}]  We vary the apocenter radius of our stellar streams from $30$--$100$~kpc, in 8 steps. The combination of apocenter radius and circularity completely describes the stream orbits in our spherical potential.

\item[{\it Circularity:}] We define stream orbit circularity, $\eta$, as the total angular momentum with respect to the angular momentum of a circular orbit with the same energy, $\eta=L/L_{circ} (E)$. We vary this parameter from eccentric to circular in 8 steps.

\item[{\it Stream Age:}] We define the age of the stream as the total time over which the mock stream generator is evolved. We set ages for our streams from $0.5$--$6.0$~Gyrs, also in 8 steps.
\end{enumerate}

 Additionally, we give each stream a random initial position and velocity direction at their assigned apocentric distance on the sphere, consistent with their prescribed orbital properties. These variations provide us with a total of $2\times8\times 8\times8=1024$ stellar streams, spanning a wide range of orbits and positions in the host halo.

\begin{table}
\caption{This table shows the property of streams varied (column 1) and their corresponding values (in column 2).}
\label{table:tab1}
\begin{ruledtabular}
\begin{tabular}{ll}
\textrm{Stellar stream properties} & \textrm{Range of values} \\
\hline
\textrm{Total stellar mass} ($M_{\odot}$) & $10^4, 10^6$ \\
\textrm{Apocenter radius (kpc)} & 30, 40, 50, 60, 70, 80, 90, 100 \\
\textrm{Age of the stream (Gyr)} & 0.5, 1, 1.5, 2, 3, 4, 5, 6 \\
\textrm{Circularity} & 0.5, 0.6, 0.7, 0.75, 0.8, 0.9, 0.95, 1.0 \\
\end{tabular}
\end{ruledtabular}
\end{table}




\subsection{Evolving streams in N-body simulations}\label{method}
Following the generation of stellar streams as described in Section~\ref{method:static}, we add them to simulations of a MW--like dark matter halo and study their evolution. Below we describe our isolated (Section~\ref{method:iso}) and merger (Section~\ref{method:merger}) simulations and detail our N-body setup (Section~\ref{method:IC}). 
Small variations in the global potential arise due to small differences between the static halo profile and the sampling of the halo distribution functions for the N-body simulations. In addition, numerical effects arise from the finite number of particles in the simulations. However, both these effects are minimal and do not affect the stream evolution (see Appendix~\ref{app:heating} for details) In both the isolated and merger runs the stellar streams do not gain additional star particles during their evolution in the merger simulations in \texttt{Gadget4}\citep{Springel+2021}, as opposed to their formation in the static halo potential in \texttt{Gala}. This results in underdensities in central regions of our stellar streams.

\subsubsection{Stellar Streams in an Isolated MW--like Halo} \label{method:iso}
We place and evolve our 1024 stellar streams in a live MW--mass dark matter halo with identical properties to our static MW--like halo and evolve them forward for 6 Gyrs (see Section~\ref{method:IC} for the detailed setup). Our isolated halo run provides a reference of stream evolution over the full time range of the simulations to compare to any evolution during a merger, and also serves as a benchmark of any numerical effects (e.g. heating) on the stellar streams in N-body simulations.

\subsubsection{Stellar Streams in a MW--LMC-like Merger} \label{method:merger}
We place the same stellar streams in the MW--like dark matter halo described in Section~\ref{method:iso}, and let it undergo a merger with a subhalo (an infalling satellite or perturber) that is 1/5th of the total mass of the MW--like halo. We explore effects of varying the pertuber mass in future work (Weerasooriya et al. in prep.). The exact setup of these simulations are described in Section~\ref{method:IC}.

In order to investigate the impact of the orbital path of the perturber on the dynamics and evolution of stellar streams, we run simulations using several distinct perturber orbits for the 1:5 merger. In Cartesian coordinates the perturber initial position is always set to $r_{sat,0}=[300,0,0]\, \text{kpc}$ (not the initial orbit apocenter), and the initial velocities are set to: $v_{sat,0}=[-100,30,0]\,\text{km} \text{s}^{-1}$ for the original orbit (corresponding to an initial circularity of $\eta=0.27$), $v_{sat,0}=[-100,45,0]\,\text{km} \text{s}^{-1}$ for the more circular orbit ($\eta=0.39$), and $v_{sat,0}=[-75,0,0]\,\text{km} \text{s}^{-1}$ for the radial orbit ($\eta=0.0$). We note that our circular orbit does not fully merge within the simulation's 6~Gyr timescale. For an even more circular orbit, the 1:5 merger would require even more time to complete in this controlled setup. A comparison of the different perturber orbits can be found in Figure~\ref{fig:r_time_satelliteorbits} in Appendix~\ref{app:orbits}.
 

\begin{figure*}
    \centering
    \centering
\includegraphics[clip=true,trim={0cm 0 0cm 0},width=.83\textwidth]{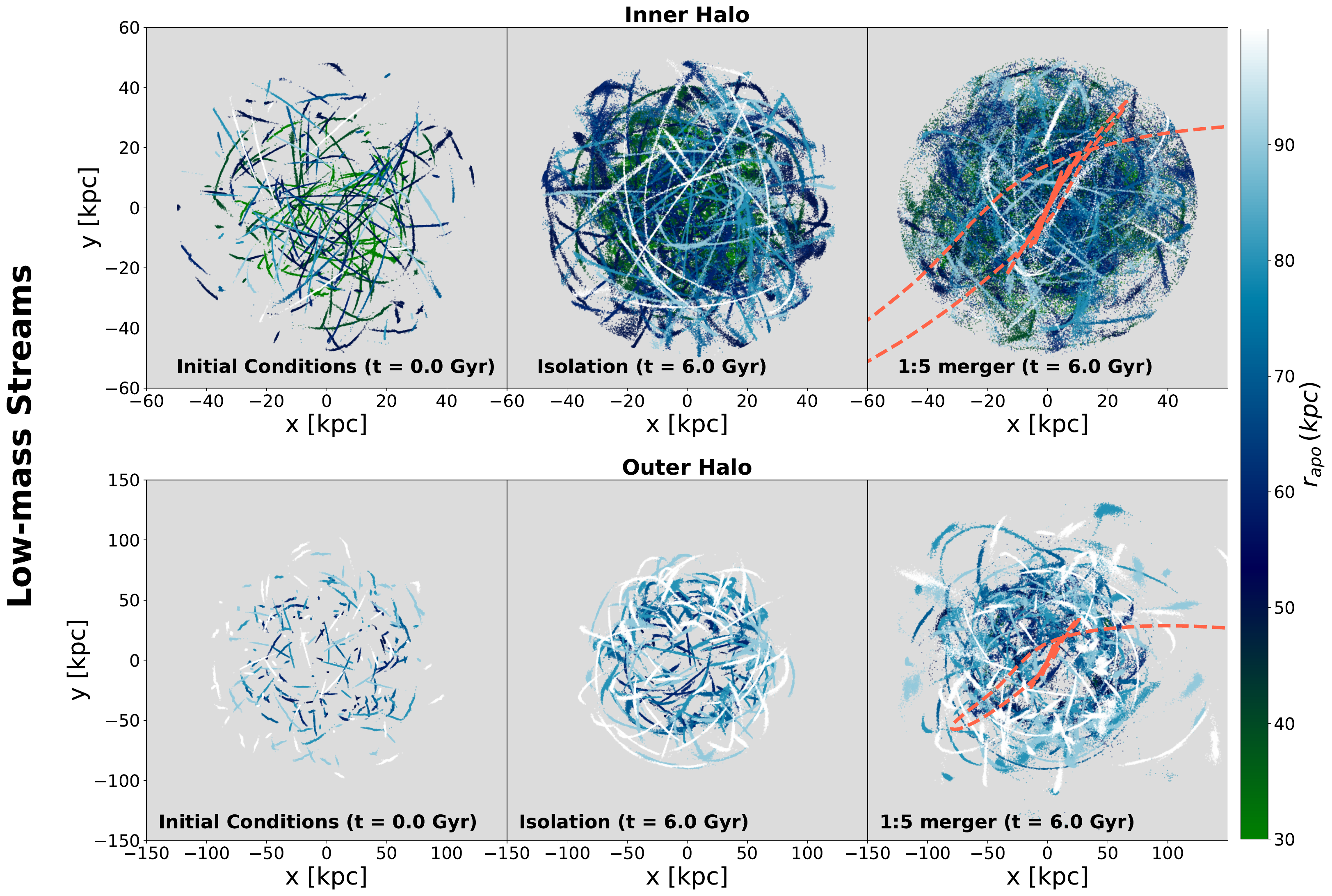}
\\
\includegraphics[clip=true,trim={0cm 0cm 0cm 0},width=.83\textwidth]{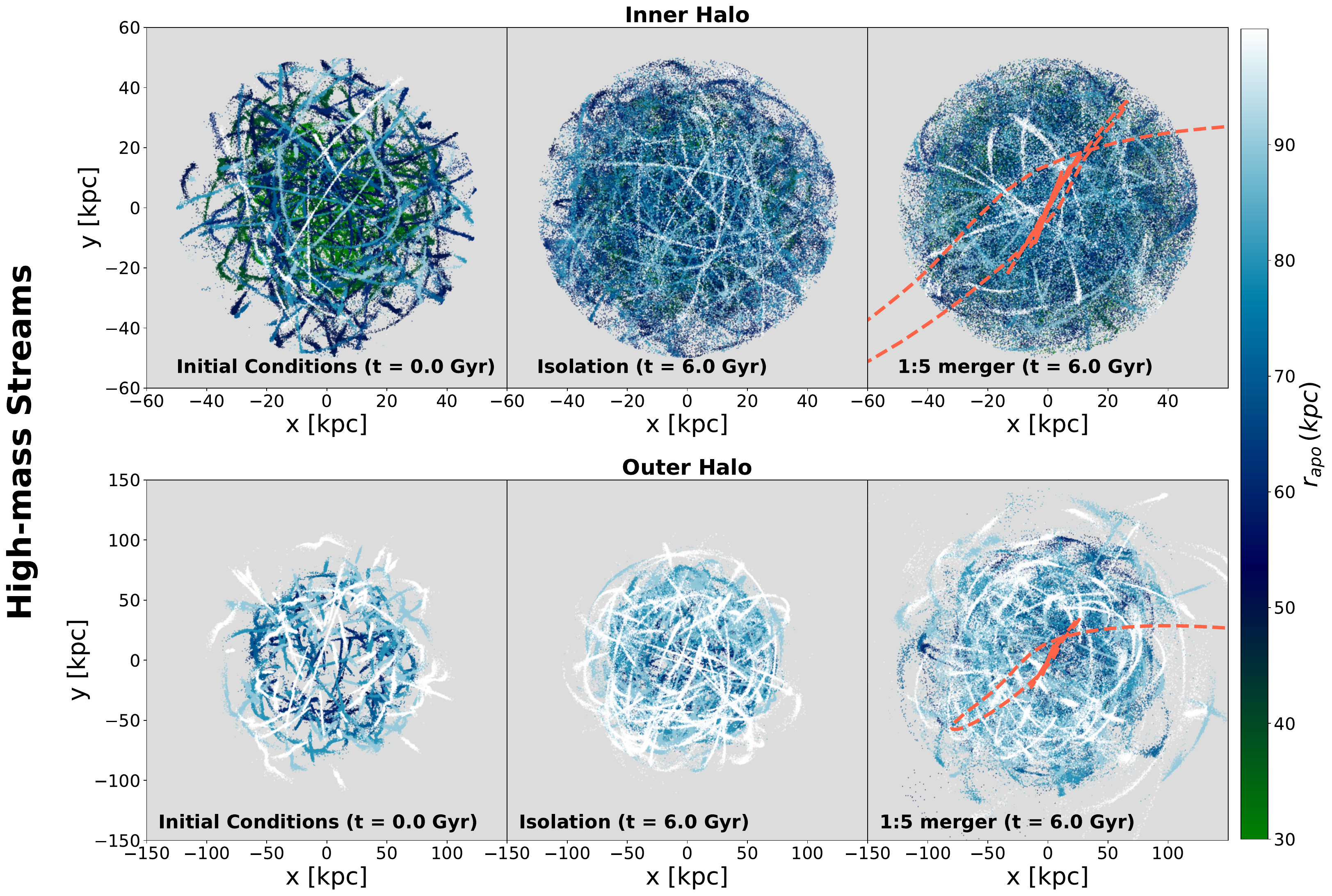}
    \caption{Initial and final positions for star particles from all stellar streams (512 low--mass ($10^4\,M_{\odot}$) and 512 high--mass ($10^6\,M_{\odot}$)) evolved for 6~Gyr in a MW--like halo, in isolation and during a 1:5 merger. Top panels (first and third row) show star particles in the inner halo ($r \leq 50,\mathrm{kpc}$) and bottom panels (second and fourth row) contain star particles in the outer halo ($r > 50,\mathrm{kpc}$). The perturber halo's orbit is shown as a dotted line, and streams are color-coded by their initial apocenter radius. Inner halo streams become more dispersed during the merger, while streams at large radii are noticeably changed in morphology and orbital parameters. high--mass streams are thicker than low--mass streams on similar orbits.}  
    \label{fig:fig1}
\end{figure*}

\subsubsection{Initial Conditions of the N-body Simulations} \label{method:IC}
In all cases, we calculate the self gravity of the particles. We evolve our set of 1024 stellar streams in an isolated N-body simulation using \texttt{Gadget4} \citep{Springel+2021}. The initial conditions for the MW potential are computed using \texttt{AGAMA} \citep{Vasiliev+2019}, with the halo parameters set based on \citet{GaravitoCamargo+2019}. This halo model is identical to the static dark matter halo in which the streams have been formed ($M_{\rm halo} = 1.57\times10^{12}\,M_{\odot}$, $a = 40.85$~kpc). The halo contains $10^7$ dark matter particles, resulting in a dark matter particle mass of $m_{\rm dm} = 1.57\times10^5\,M_{\odot}$. The perturber has a total mass and particle number set by the merger ratio ensuring an identical particle mass for the host and perturbers halos. For the mergers described in this paper, with a merger rate of 1:5, the mass of the perturber is $3.14\times 10^{11}\,M_{\odot}$, which is approximately equivalent to the mass of the Large Magellanic Cloud (LMC) \cite{GaravitoCamargo+2019}. With $1000$ particles per stream, the stream particle masses are $m_* = 10\,M_{\odot}$ and $m_* = 10^3\,M_{\odot}$ for the low and high--mass streams respectively.

In the merger simulations we ensure that the dark matter particle masses of the host halo and perturber are identical, thus the merger ratio between the total halo masses also provides the ratio between the number of particles in each halo. 

In all figures and results all particles are re-centered to the iteratively calculated center-of-mass of the host halo. We determine the center of both the parent halo and the perturber using the iterative shrinking sphere method as outlined by \citep{Power+2003}, reducing the sphere size by $2.5\%$ with each successive step.

\begin{figure*}[t]
    \centering
    \includegraphics[clip=true,trim={0cm 0cm 0cm 0},width=\textwidth]{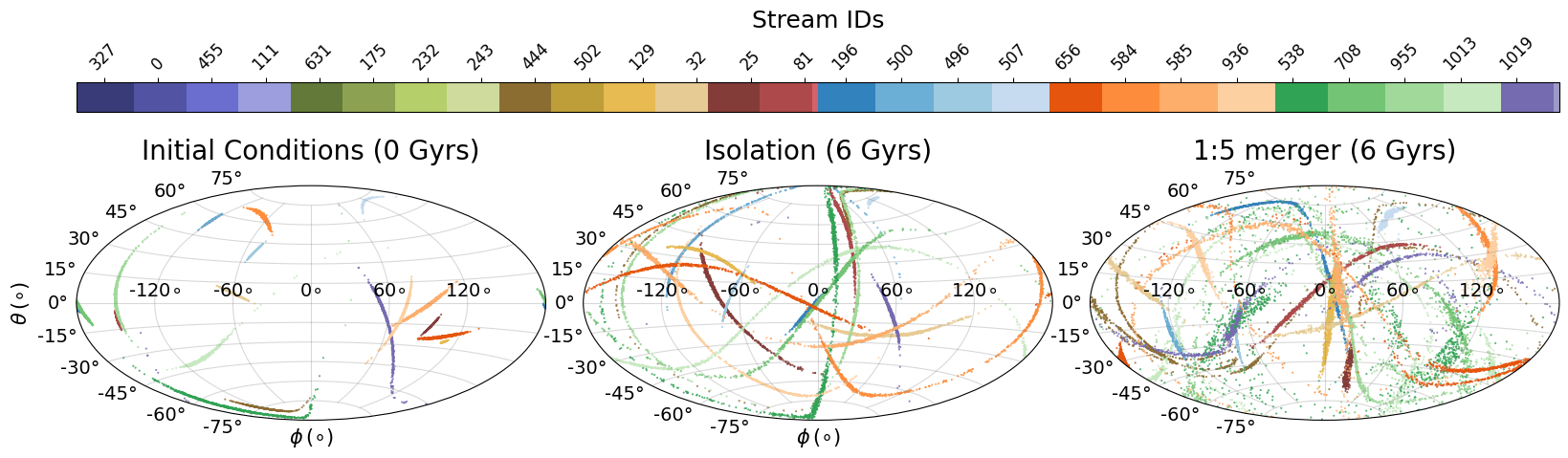}
    \includegraphics[clip=true,trim={0cm 0 0cm 0},width=\textwidth]{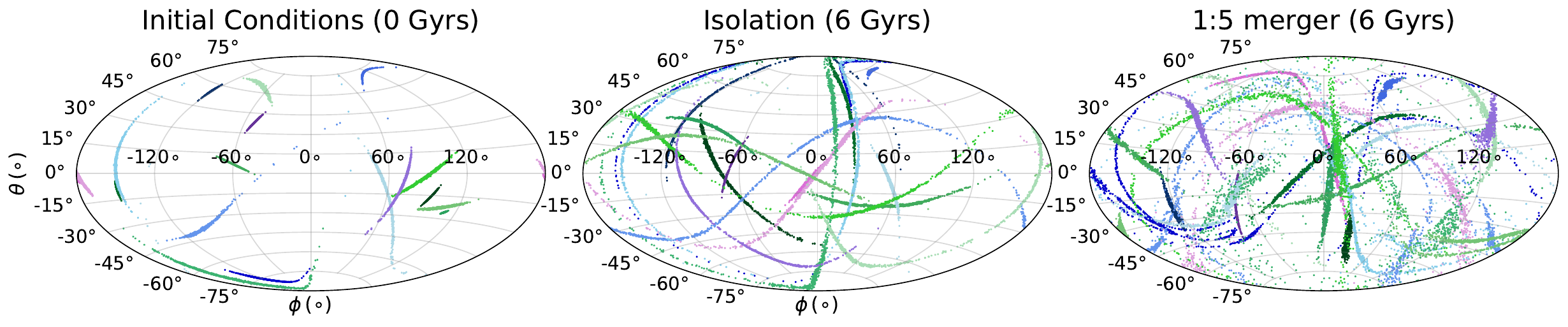}
    \caption{A subset of 18 stellar streams in an Aitoff projection at 0 Gyrs (left), evolved in isolation after 6 Gyrs (middle), and in our 1:5 fiducial merger after 6 Gyrs (right). Top panel: the 18 selected streams colored by stream ID following the color coding of Figure~\ref{fig:movie}. Bottom panel: The identical subset of 18 stellar streams grouped and colored according to their orbital properties. The first groups contains older ($5$--$6$~Gyr) streams on outer ($90$--$100$~kpc), less circular orbits ($\eta = 0.7$--$0.8$) and is shown in shades of blue, and the second group exists of younger ($0.5$--$3$~Gyr) streams on inner ($30$--$50$~kpc), close-to-circular orbits ($\eta = 0.9$--$1.0$) and is shown in shades of green. Added to that are two older streams ($4$ and $5$ Gyr) on circular orbits ($\eta = 1.0$) with large apocenter radii ($r_a = 90$--$100$ kpc) shown in shades of purple, and two young streams ($0.5$ Gyr) on less circular ($\eta = 0.75$), intermediate ($r_a = 60$ kpc) orbits with opposite angular momentum directions, shown in shades of pink. Initial and final properties for each stream are described in Table~\ref{tab:tab3} and each stream is shaded by its order in Table~\ref{tab:tab3}, from dark to light, and the colors are identical to those in Figure~\ref{fig:fig3}. Almost all streams show significant perturbation in the 1:5 merger case compared to evolution in isolation, most notably, orbital planes have tilted and streams are more dispersed. The selected streams shown here can also be found in Figure~\ref{fig:movie} and in Figure~\ref{fig:fig3} in orbital coordinates.}
    \label{fig:selected}
\end{figure*}




\section{Stellar Streams In a MW -- LMC-like Mass Ratio Merger}\label{sec:results}

In our fiducial merger case --- the merger of an approximately MW--mass halo with a perturber that is $20\%$ of the main halo's mass and can be thought of as Large Magellanic Cloud-like, on an intermediate orbit (with a circularity of 0.27 and a merger timescale of ${\sim}3$~Gyr, see also Appendix~\ref{app:orbits} for a description of the perturber orbit) --- all stellar streams are affected in some way, but the effects show a huge range.

In this Section we discuss the main evolutionary trends of the stellar streams in the merger versus the isolated simulations. While we cannot discuss every possible effect of merger on stellar streams, it is still important to understand how each stream could be affected individually as well as the whole population. 

\subsection{Visual Impression of all streams}
Figure \ref{fig:fig1} shows all streams (the low--mass streams in the top two rows, the high--mass streams in the bottom two rows), evolved in isolation or during our 1:5 mass ratio fiducial merger: showing the initial conditions (ICs, left panels), for all low--mass streams (left panels), along with the streams evolved for 6 Gyrs in their host halo in isolation (middle panels), and evolved during a merger (4.5 Gyrs after the merger, right panels). The streams are color-coded based on their initial apocenter radius and all individual star particles are separated by distance from the halo center: into the inner halo ($\leq 50\,\mathrm{kpc}$; top and third row) and the outer halo ($> 50\,\mathrm{kpc}$; second and last row). While during the 6 Gyr all streams have grown in length and span a significant part of their orbit, or even wrap around, visual comparison immediately highlights the effect of the merger on the stellar streams: the streams evolved during a merger appear significantly disturbed in addition to growing in length and width. It is also apparent that some have moved to orbits with larger apocenters (evident when comparing the panels for the outer halo). Streams located closer toward the center of the host halo appear more dispersed, significantly gaining in thickness compared to their evolution in isolation, for some up to the point of losing a stream-like morphology. Most stellar streams in the inner halo after 6 Gyrs in a 1:5 merger cannot be identified distinctively except for those streams with high initial apocenters (blue-white colors). The streams in the outer halo have largely retained their stream-like appearances, but still show significant morphological evolution and changes in orbits moving to outskirts of the halo. However, the detailed evolution of morphological and orbital properties of stellar streams during complete merger events remains unexplored.
We will highlight some of the streams with striking morphologies in Section \ref{sec:individual}. Figure~\ref{fig:fig1} also highlights differences between the high--mass and low--mass streams:  high--mass streams are broader inherently but also appear more disturbed and dispersed in both the inner and outer halo than their lower-mass counterparts. The increase in thickness of all stellar streams and a quantitative comparison of effects for low- and high--mass streams is discussed in Section~\ref{sec:15stat} and shown in Figure~\ref{fig:fig5}.

 In this section we describe the evolution of the stellar streams as a result of the merger in a statistical sense (Section~\ref{sec:15stat}), with Section~\ref{sec:individual} going into individual streams. However, we will first discuss some aspects of the morphological and orbital evolution of streams through describing that of a few selections of streams with similar initial properties.

\subsection{Characterizing evolution with a select sample}
\label{selection}

We select 18 example stellar streams and group them as follows:

\begin{itemize}
\item Group 1: older streams on less circular orbits with large apocenter radii (shown in blue); 

\item Group 2: young streams on close-to-circular orbits with small apocenters (shown in green); 

\item Group 3: two streams that physically cross these groups (older streams on circular initial orbits with large apocenters, shown in dark blue)

\item Group 4: two streams on close to identical orbits but with opposite orientation (one prograde, one retrograde with respect to the satellite's orbit, shown in shades of purple).
\end{itemize}

A complete list of selected example streams and their initial and final properties are listed in Table~\ref{tab:tab3}.

\subsubsection{Broad Characteristics}

Figure~\ref{fig:selected} shows the same 18 streams in an Aitoff projection at 0 Gyrs (left panel), evolved in isolation after 6 Gyrs (mid panel), and in a 1:5 merger after 6 Gyrs (right panel).
Figure~\ref{fig:selected} further highlights how streams grow in length and get perturbed away from their original orbital plane: all streams have become thicker and less well-defined in the merger case. While comparing the initial conditions (left) and isolated halo results (middle) finds streams on the same orbits (of course often not the same phase), a comparison with the 1:5 merger results show that almost all streams end up on completely different orbits irrespective of circularity. For example the two streams with identical properties but opposite angular momentum vectors are right on top of each other in the top middle panel (purple streams), but have different orbits.

\subsubsection{Selecting streams individually}

\begin{figure*}[p]
\centering
    \includegraphics[width=0.45\textwidth]{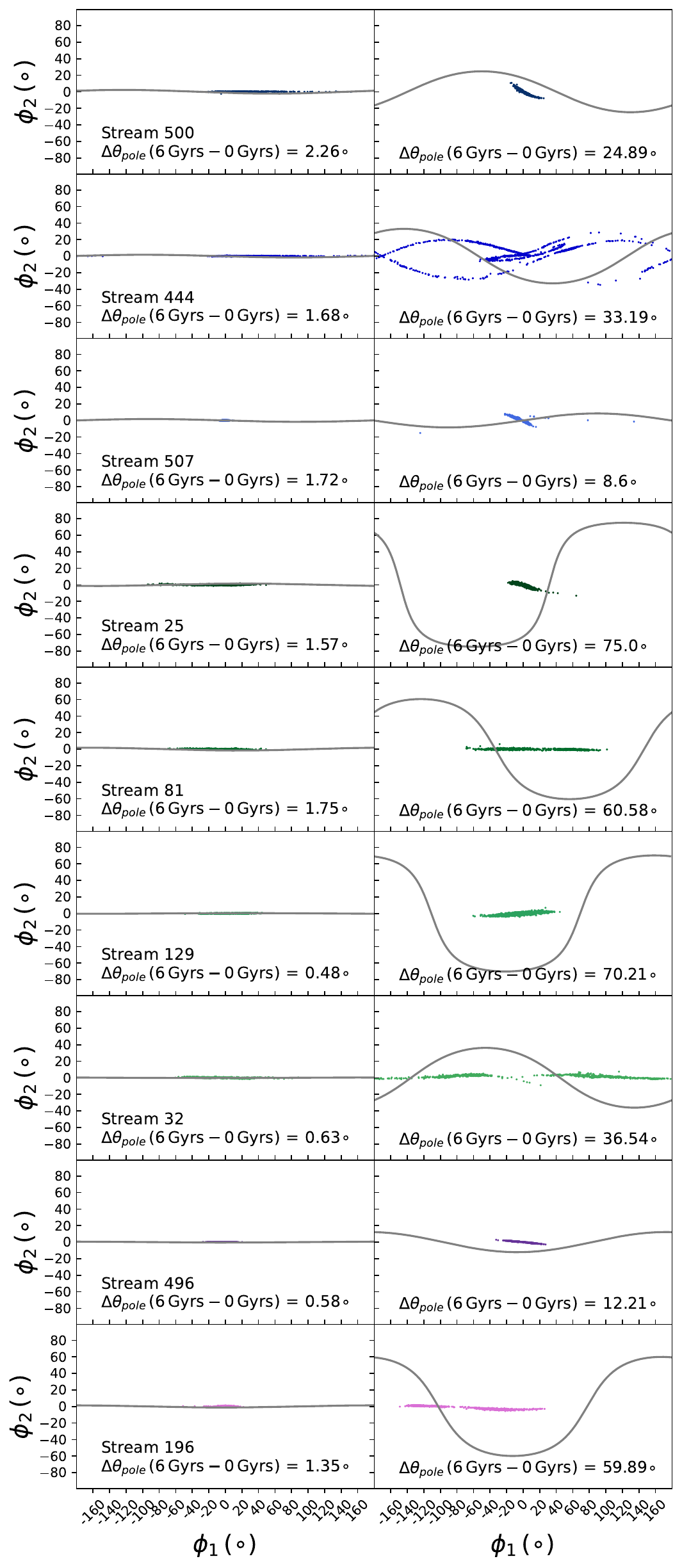}
\hfill
\includegraphics[width=0.45\textwidth]{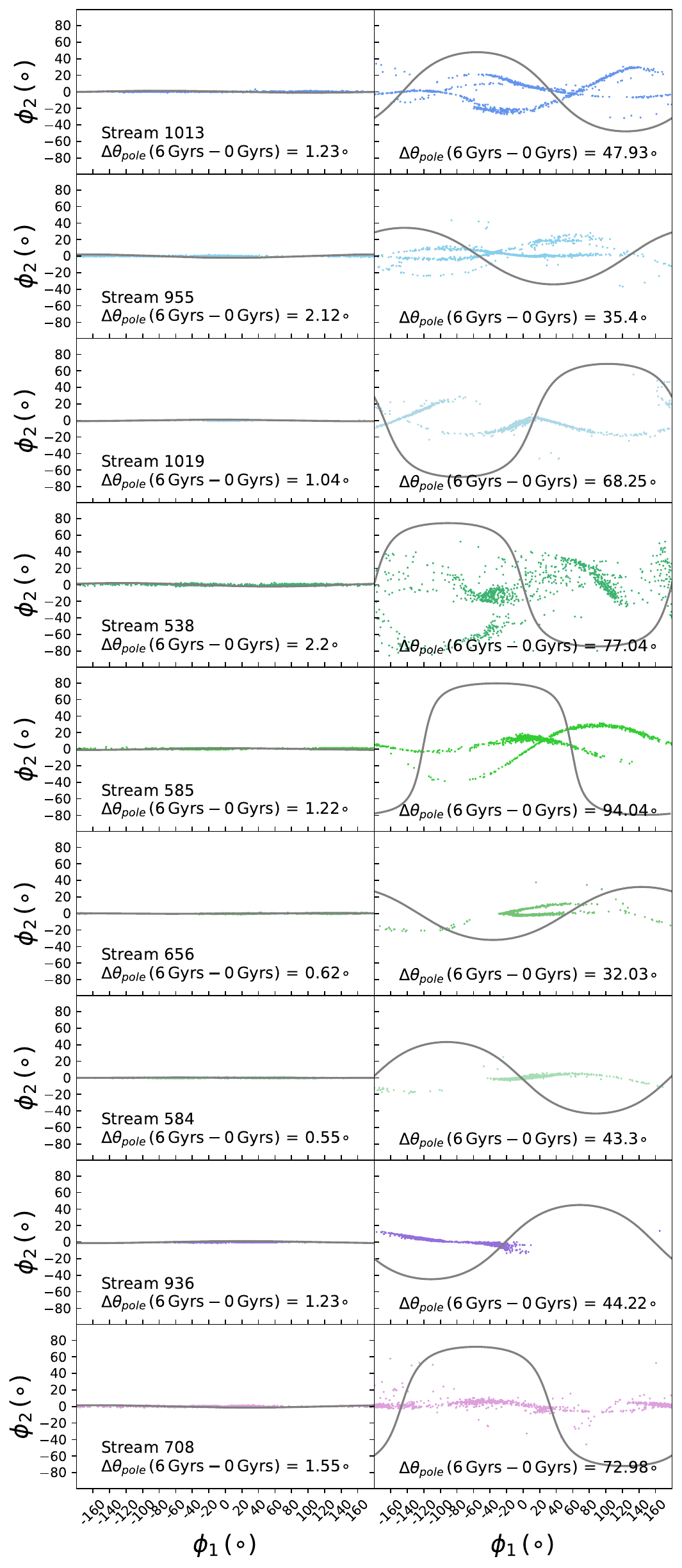}
    \caption{The same streams as in Figure~\ref{fig:selected} shown in their orbital plane coordinates, evolved in isolation (first and third columns) and during the 1:5 merger (second and fourth columns). Streams are shown in the same order as in Table~\ref{tab:tab3} with the left pair of columns showing low--mass streams and the right pair of panels high--mass streams. All streams are colored by their respective groups identical to Figure~\ref{fig:selected}. The orginal (0 Gyr) orbit of every stream is overplotted in the stream's new coordinated frame in grey. All streams are well behaved when evolved in isolation and predominantly grow in length. In the merger case there are a number of streams with bifurcation and/or separations into multiple stream--like structures, as well as an increase in stream thickness and re-orientations of the orbital plane. More extreme examples of these effects are discussed in Section~\ref{sec:individual}.}
    \label{fig:fig3}
\end{figure*}

Figure~\ref{fig:fig3} shows these 18 streams in great circle coordinates, where each stream is oriented on its own orbital plane: along, $\phi_1$, and perpendicular, $\phi_2$, to the plane. To determine a stream's orbital plane we calculate the location of the orbital pole ---the direction of the angular momentum vector--- for all star particles in a stream, take the median of those locations, and define the stream's orbital plane as the great circle corresponding to that median pole. \\

We calculate the dispersion of the orbital pole directions and find it is smaller than 2$^{\circ}$ for the isolated streams. If stream stars get dispersed or moved to a different orbit they will be tilted away from the median orbital plane. See Appendix \ref{app:heating1} for a resolution study.


We note the following points with respect to Figure~\ref{fig:fig3}:
\begin{itemize}
    \item \textbf{Isolated streams remain thin.} Consider first the streams in the isolated case (left hand columns). In an unperturbed, spherical potential, such as in the isolated halo simulations, streams start to overlap in great circle coordinates and in sky projections when they span beyond the full 360 degrees in $\phi_1$, forming multiple wraps around the host halo (e.g. streams 1013, 708, 955, 538, 585).  At the same time, all these streams remain thin in $\phi_2$. We calculate the dispersion of the orbital pole directions and find it is smaller than 2$^{\circ}$ for the isolated streams. As noted in methods, because we do not add new stream star particles while the streams evolve in the N-body simulations, underdensities develop in the central regions of the streams.
    
    \item \textbf{Streams can be either longer or shorter in the perturbed case.} Over the 6 Gyr of evolution all streams grow in length, as expected, both in the isolated and merger runs. In some cases (e.g. streams 507, 32, 196 in the right column) perturbed streams become noticeably longer than the isolated stream, but in other cases this appears to be the other way around (e.g. streams 500, 25). 
    
    \item \textbf{Stars in perturbed streams can significantly deviate from the stream's median orbital plane, leading to bifurcations and oscillatory substructures (the ``waves") where one part of the stream is on a different orbit than other parts of the stream.} The waves also show that these streams in fact exist out of multiple wraps. These are more apparent in the higher mass streams for the subset of streams shown (see, e.g. stream 444 left column, or streams 1013, 955, 1019, 538, 585, 656 in the right column). We further test the robustness of this result in a resolution study in the Appendix \ref{app:heating1}. If stream stars are dispersed or moved to a different orbit, they will be tilted away from the median orbital plane. The same is not the case in non-spherical potentials: in a static non-spherical potential because of precession and in an evolving potential because of ongoing changes to the potential along the stream and orbit. In our merger models, both the incoming perturber and the reaction to this from the main halo will continually change the potential, locally and globally. 

    \item Additionally, streams perturbed by a merger can develop a gradient along the orbits, where stream stars are offset perpendicular to the orbital plane. These streams appear cohesive but tilted in great circle coordinates (see e.g. the first, fourth and fifth stream in the left column, or the last stream in the right column).

    \item \textbf{Entire streams can evolve significantly in their orientation due to the interaction:} Annotated on each panel is the change in the median pole with respect to the initial value. While streams can slightly change their orbital plane in isolation, this is negligible (a few degrees at most). All streams in the merger simulation see significant changes in their mean orbital pole direction: only 3 out of 18 streams change their median orbital plane by $< 30^{\circ}$ (these are streams at large apocenters on circular or circular orbits), and many have $> 60^{\circ}$ going up to $94^{\circ}$. \emph{These values highlight that the position on the sky should change significantly during a merger for a large number of streams, even when the streams still appear thin and relatively smooth.}

    \item \textbf{A number of, in this selection, mostly low--mass streams appear barely perturbed in their orbital frame}: they are a bit thicker, and have gained a very slight tilt in the great circle coordinates, even though their orbital plane has rotated (e.g. the sixth, seventh, and ninth streams in the left column). These three streams are on close-to circular orbits and are relatively short streams all throughout the simulation.
    
\end{itemize}

\begin{table*}
\newlength\q
\setlength\q{1.55cm}
    \centering
    \begin{tabular}{|ccccc|cccc|cccccc|}
    \hline
    \multicolumn{15}{|c|}{low--mass streams}\\
    \hline
    \multicolumn{5}{|c|}{Initial} & \multicolumn{4}{c|}{Final Isolation} & \multicolumn{6}{c|}{Final Merger}\\
    \hline
     gr. & ID & $r_{apo}$&Age&$\eta$&$\Delta \theta_{\rm pole}$&$\sigma_{\phi_2}$&$\Delta E_{\rm med}$&$\Delta L_{med}$ &$\Delta \theta_{\rm pole}$&$\sigma_{\phi_2}$&$\Delta E_{\rm med}$&$\Delta L_{med}$&$d_{\rm closest}$&$t_{\rm closest}$\\
     & &(kpc)&(Gyr)& &(deg)&(deg)& $(\frac{km}{s})^2$ &($\frac{kpc\,km}{s}$) &(deg)&(deg)&$(\frac{km}{s})^2$&($\frac{kpc\,km}{s}$) &(kpc)&(Gyr)\\
    \hline
    \hline
       1 & 500 & 100 & 5.0 & 0.75 &2.3 & 0.3  &  34  &  55 & 24.9 & 1.9  &  -10890  &  -2342 & 30.2 & 1.6\\
        
        1& 444 & 90 & 6.0 & 0.75  &1.7 & 0.5  &  -311  &  -8 &33.2 & 2.6  &  6227  &  -945 & 13.6 & 1.6\\
        
       1& 507 & 100 & 6.0 & 0.8  &1.7 & 0.4  &  -642  &  -128 & 8.6 & 0.7  &  7256  &  1084 & 12.8 & 2.5\\

        2 & 25 & 30 & 2.0 & 0.95  &1.6 & 0.4  &  -646  &  -104 &75.0 & 2.2  &  286  &  1284  &0.5 & 2.9\\
        
       2 &  81 & 40 & 1.5 & 0.95  &1.8 & 0.3  &  63  &  68 &60.6 & 2.8  &  6954  &  5368 & 27.5 & 2.9\\
        
         2 & 129 & 50 & 0.5 & 0.95  &0.5 & 0.1  &  -103  &  -35 &70.2 & 1.0  &  2597  &  2984  &69.2 & 2.4\\
         
        2& 32 & 30 & 3.0 & 1.0  &0.6 & 0.3  &  278  &  -6 &36.5 & 9.9  &  -19476  &  -795   &11.8 & 2.5\\
        
        3& 496  & 100 & 5.0 & 1.0  &0.6 & 0.2  &  377  &  114 &12.2 & 2.7  &  1856  &  2138 & 21.8 & 2.4\\
        
        4 & 196 & 60 & 0.5 & 0.75  &1.4 & 0.5  &  77  &  2 &59.9 & 1.8  &  -9379  &  -558 &  12.4 & 2.8\\
    \hline
    \multicolumn{15}{|c|}{high--mass streams}\\
    \hline
    \multicolumn{5}{|c|}{Initial} & \multicolumn{4}{c|}{Final Isolation} & \multicolumn{6}{c|}{Final Merger}\\
    \hline
     gr. & ID & $r_{apo}$&Age&$\eta$&$\Delta \theta_{pole}$&$\sigma_{\phi_2}$&$\Delta E_{med}$&$\Delta L_{med}$&$\Delta \theta_{pole}$&$\sigma_{\phi_2}$&$\Delta E_{med}$&$\Delta L_{med}$&$d_{\rm closest}$&$t_{\rm closest}$\\
      & &(kpc)&(Gyr)& &(deg)&(deg)&$(\frac{km}{s})^2$&($\frac{kpc\,km}{s}$) &(deg)&(deg)&$(\frac{km}{s}^2$&($\frac{kpc\,km}{s}$) &(kpc)&(Gyr)\\
    \hline
    \hline
       1 & 1013 & 100 & 5.0 & 0.7  &1.2 & 0.4  &  688  &  226 &47.9 & 14.4  &  -5241  &  -192 & 9.2 & 2.8\\
       
       1 & 955 & 90 & 6.0 & 0.8  &2.1 & 0.4  &  561  &  56 &35.4 & 8.4  &  12344  &  3181 & 12.7 & 1.5\\
       
       1 & 1019 & 100 & 6.0 & 0.8  &1.0 & 0.3  &  498  &  311 &68.3 & 11.1  &  -8664  &  -1689 & 26.4 & 1.5\\
       
      2 & 538 & 30 & 2.0 & 0.9 & 2.2& 0.8  &  83  &  32 & 77.0 & 27.5  &  -7592  &  -1764 & 3.4 & 2.9\\
       
       2 & 585 & 40 & 1.0 & 0.95 &1.2 & 0.5  &  561  &  56 & 94.0 & 11.9  &  -3500  &  186 & 12.6 & 2.5\\
       
       2 & 656 & 50 & 1.5 & 1.0 &0.6 & 0.4  &  221  &  77 & 32.0 & 5.3  &  7881  &  2685 & 2.2 & 1.6\\
       
       2 & 584 & 40 & 1.0 & 1.0 &0.6 & 0.5  &  782  &  136 & 43.3 & 3.3  &  8546  &  142 & 4.0 & 2.4\\
       
       3 & 936 & 90 & 4.0 & 1.0 &1.2 & 0.3  &  -366  &  -152 & 44.2 & 14.4  &  1200  &  2157 & 66.2 & 1.5\\
               
       4 & 708 & 60 & 0.5 & 0.75  &1.6 & 0.6  &  -442  &  27 &73.0 & 6.6  &  -4015  &  677 & 0.4 & 3.0\\
        \hline
    \hline
    \end{tabular}   
    \caption{Properties of 18 selected streams, from two groups with comparable initial apocenter radii, circularity, and age (group 1 and group 2), two streams that have properties across these groups (group 3), and two streams with identical initial conditions but different masses and opposite angular momentum directions (group 4). The top 9 streams are low--mass streams while the second set are high--mass streams. Within groups, streams are ordered by circularity and the full order with the low (high)-mass streams is identical to that in the left (right) column of Figure~\ref{fig:fig3}. Column 1: group number; Column 2: Stream ID. Column 3--5: initial properties: apocenter radius, stream age, and orbit circularity; Column 6--9: final stream properties after 6 Gyr evolution in an isolated halo: change in orientation of the orbital plane, $\Delta \theta_{pole}$, stream thickness, $\sigma_{\phi_2}$, change in median Energy, $\Delta E_{med}$, and change in median angular momentum $\Delta L_{med}$; Column 10--15: Final properties after 6 Gyr with a 1:5 merger: the same properties as for the isolated case, as well as the closest distance between any stream star particle and the center of mass of the perturber, before the perturber has merged into the center of the host halo, $d_{\rm closest}$, and the time at which this occurs, $t_{\rm closest}$; For reference, the first pericenter passage is at approximately $t{\sim}1.5\, {\rm Gyr}$, second pericenter at $t{\sim}2.5\, {\rm Gyr}$ and after third pericenter at $t{\sim}3\, {\rm Gyr}$ we consider the perturber as merged in (see Appendix~\ref{app:orbits} for the perturber orbit).\label{tab:tab3}}
\end{table*}

\subsection{Quantifying stream evolution}
\label{sec:15stat}

We use the results from the prior sections as guidance on how to quantify trends in the morphological evolution of our full population of streams. The complexity in Figure \ref{fig:fig3} underlines the need to take the individual complexity of each stellar stream into account when considering summary statistics for the population.
We focus our quantified description of the stream evolution on the increased width, or dispersion, perpendicular to the stream track, $\Delta (\sigma \phi_2)$, changes in the orientation of the orbital plane of each stream, $\Delta \theta_{pole}$, and changes in the median energy and angular momentum of each stream. We note that an increase in width may be due either to larger dispersion in the stream (see e.g. Stream 496, the 7th stream in the left column of Figure~\ref{fig:fig3}) or to systematic differences in local properties at different points along the stream, for example a difference in orbital poles along the stream (e.g. Stream 129, the 1st stream in the left column of Figure~\ref{fig:fig3}), bifurcations or stream splitting (Stream 25, the 3rd stream in the left column of Figure~\ref{fig:fig3}), or a combination of these (Stream 538, the 5th stream in the right column of Figure~\ref{fig:fig3}). We have also considered the dispersion in orbital pole direction within streams as a separate parameter, but find it does not provide additional information. Values for these parameters are provided for the selected streams from Section~\ref{selection} in Table~\ref{tab:tab3}, describing their evolution in the isolated and merger simulations.

\begin{figure*}[t]
    \centering
    \includegraphics[width=\textwidth]{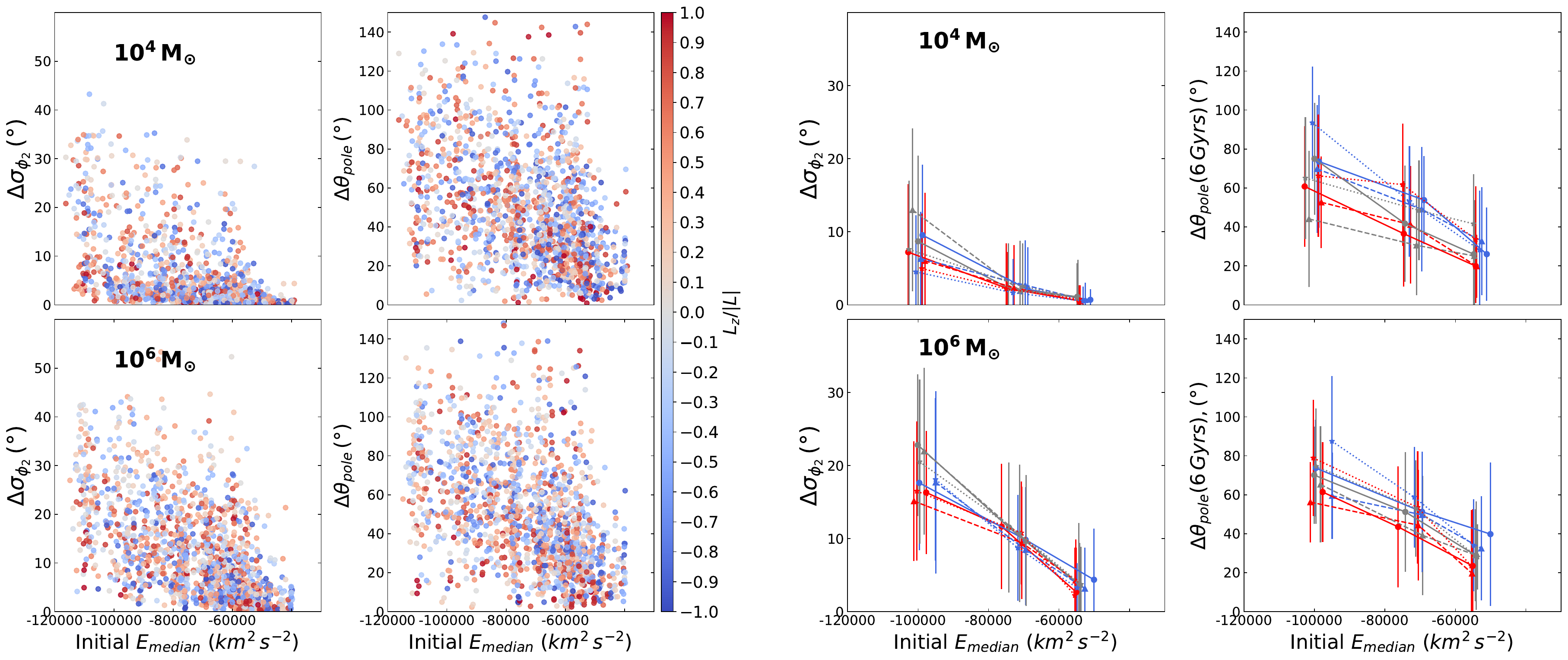}
    \caption{The change in dispersion of $\phi_2$ for each stream between 0 and 6 Gyrs (1st column) and the change in pole between 0 and 6 Gyrs (second column) with respect to their initial median energy with top rows showing low--mass streams bottom panels high--mass streams. Each stream is colored by their orbital orientation with respect to the $xy$-plane, $L_z/|L|$. Change in poles shows the angle by which the orbital plane rotated. The rightmost panels show the same data as in the first and second columns, but separating trends for streams on prograde ($L_z/|L| > 0.5$; red), retrograde ($L_z/|L| < -0.5$; blue), and in-between ($-0.3 <L_z/|L| < 0.3$, polar; grey) orbits in energy bins. Solid, dashed, and dotted lines show the original, circular, and radial perturber orbits respectively. Points show the median energy and median pole change for all streams in the bin with error bars of one standard deviation. There is a clear inverse trend of streams becoming more dispersed and changing their orbital planes by larger angles for lower energy, i.e. orbits with smaller apocenters and larger eccentricity become significantly more dispersed and diverted. Overall, high--mass streams become more dispersed in $\phi_2$, and thus more strongly increase in stream width, than low--mass streams. Low-- and high--mass streams show similar trends in pole angle changes, and the stacked data for orbital orientation suggest a possible small effect of orientation on rotations of the orbital plane.}
    \label{fig:fig5}
\end{figure*}

\begin{figure*}
    \centering
            \includegraphics[clip=true,trim={2.5cm 0.3cm 10cm 2cm},width=\textwidth]{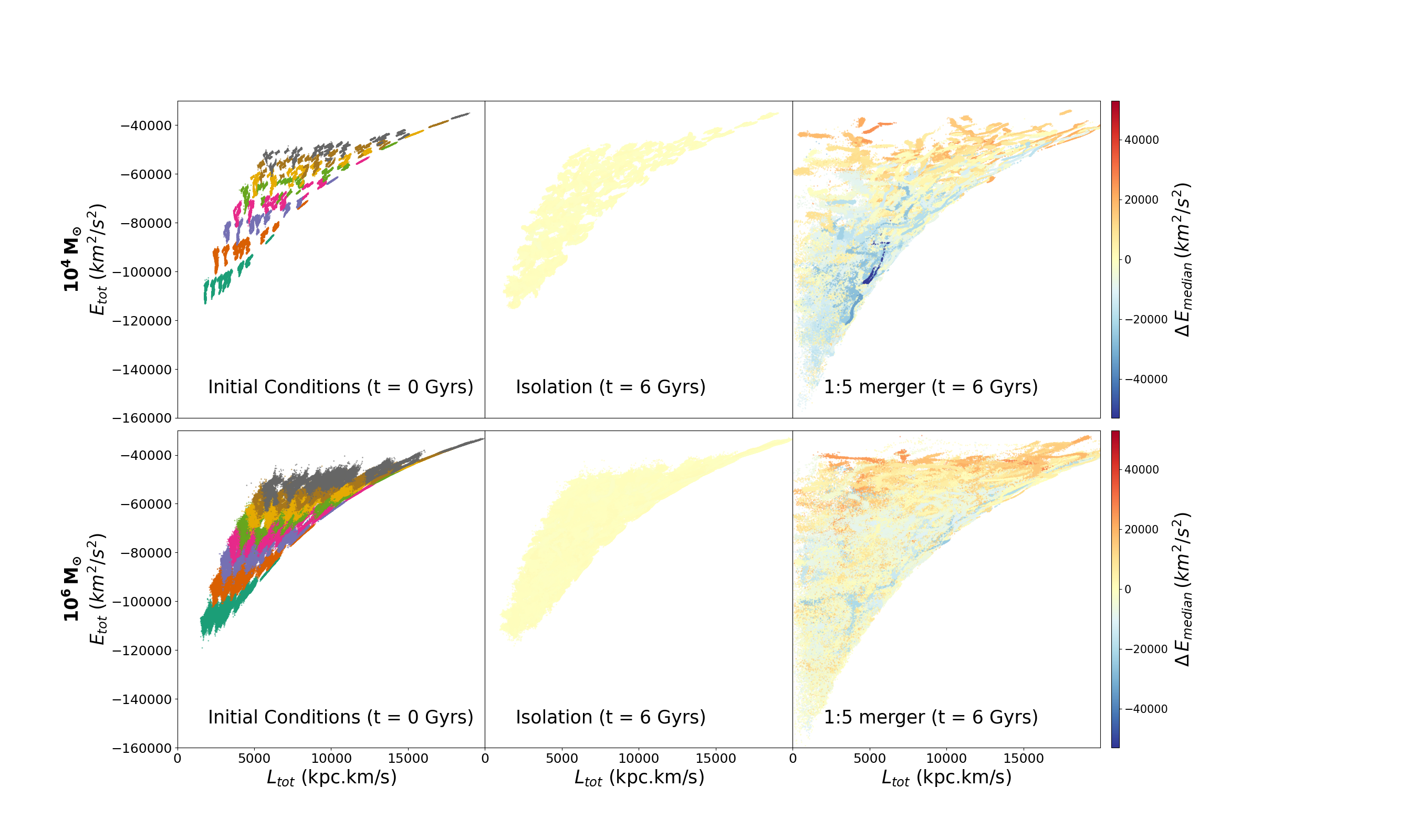}
       
\caption{
Total energy vs. total angular momentum of stream particles, initially, colored by their initial apocenters (from $r_{\rm apo} = 30$~kpc, green, to $r_{\rm apo} = 100$~kpc, gray; left panels), streams evolved in isolation at 6 Gyrs (middle panels), and streams evolved in our fiducial merger at 6 Gyrs (right panels). Both middle and right panels are colored by the change in median energy of each stream between 0 and 6 Gyrs. Top and bottom panels show $10^4\,M_{\odot}$ and $10^6\,M_{\odot}$ stellar streams respectively. The median energies of streams have minimal displacement when evolved in isolation. Streams with the lowest $L_{tot}$ and $E_{tot}$ get dispersed more during the merger, while those with high $L_{tot}$ and $E_{tot}$ appear to stay more coherent. The stream visible as the dark blue feature in the upper right-hand panel is the stream that loses the most energy in this simulation; we explore this special case in Section \ref{sec:individual}.}
\label{fig:Etot_Ltot}
\end{figure*}
Figure~\ref{fig:fig5} shows trends in the change in dispersion of $\phi_2$ and the change in orbital plane for streams in all our three merger simulations (the fiducial, circular, and radial perturber orbits), separated by stream mass. High--mass streams have a higher dispersion initially, but also gain more dispersion during the merger (even relative to their counterparts in the isolated simulation). Streams on lower energy orbits get much more dispersed for both low- and high--mass streams, although the trend is more prominent for high--mass streams. The spread in pole angle changes is similar for low--mass and high--mass streams and shows a stronger trend with orbital energy. For the streams evolved in isolation the change in dispersion is very low with a maximum of 1 degree, and the angular change in median orbital pole direction peaks around 1 degree and is always below 5 degrees, both lower than almost all of the streams evolved during a merger.

All the points are colored by the orientation of their orbit, $L_z/|L|$, which is also relative to the orbit of the perturber which orbits in the $xy$-plane. The points suggest only a small possible dependence on orientation of the stream with respect to the perturber. When separating the prograde ($L_z/|L| > 0.5$), retrograde ($L_z/|L| < -0.5$), and in-between ($-0.5 < L_z/|L| < 0.5$) streams in energy bins there is a suggestion of a small but not significant difference for the low--mass streams in the change in orbital pole, and even less strongly for the high--mass streams. There is no correlation of stream thickness with orientation of the stream orbit.

\begin{figure*}[t]
    \centering
    \includegraphics[width=0.8\textwidth]{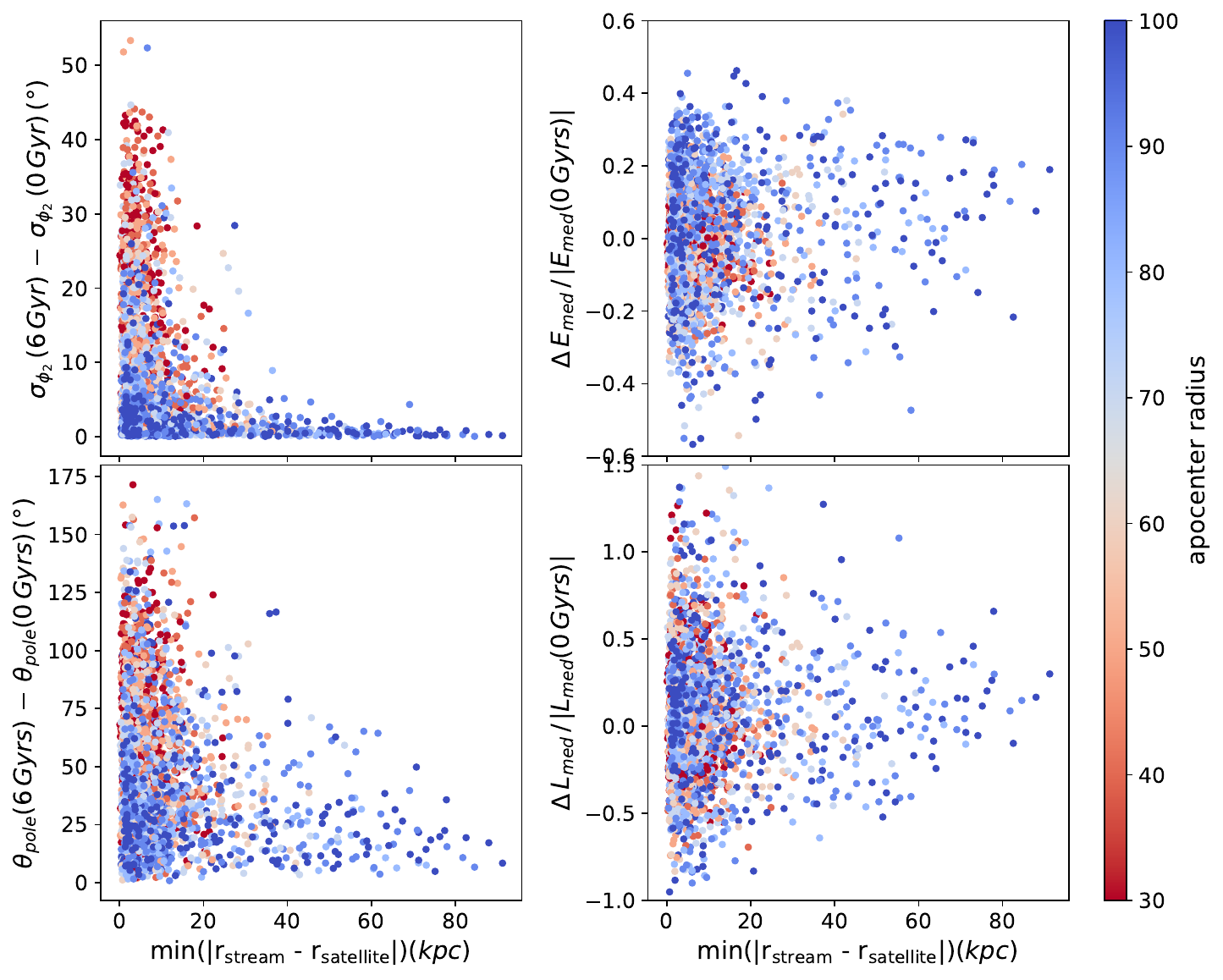}
    \caption{Distance of closest approach of \emph{any} star particle in each stream against increased dispersion perpendicular to the stream track (top left), rotation of the stream orbital plane (bottom left), relative change in median stream energy (top right), and relative change in median stream angular momentum (bottom right), color-coded by the initial apocenter radius. Close approaches of any stream star with the perturber are clearly correlated with changes in the orbital plane and dispersion, while we see somewhat weaker correlations with the relative change in median energy and the relative change in median angular momentum. Note that we include streams from perturbers with different orbits in this figure.
    }
    \label{fig:closest}
\end{figure*}

We explore the evolution of energy and angular momentum that are close to being conserved for stream stars in the isolated simulations (see also Appendix~\ref{app:heating}), Finding that both the energy and the angular momentum of streams can either increase or decrease by a small amount or a significant fraction of their initial values. Figure~\ref{fig:Etot_Ltot} shows the total energy and angular momentum of all stream star particles (particles in low--mass streams in the top row, and particles in high--mass streams in the bottom row), initially, after 6 Gyr of evolution in an isolated halo, and after 6 Gyr of evolution in our fiducial 1:5 merger. For the initial conditions the stream stars are colored by their apocenter radii (according to the same color scheme as in Figures~\ref{fig:selected} and \ref{fig:fig3}), while the isolated and merger panels are colored by the change in the median energy for each stellar stream. The extremely small change in total energy for the isolated simulation and the identical distributions between the initial conditions and the isolated final values validates the differences we see in the merger runs. The overall energy -- angular momentum space covered by all stream particles is clearly much larger after evolution during a merger than in isolation. This is not completely surprising as the perturber adds mass to the halo and thus particles can move into a deeper potential well, which allows for energy loss (when assuming no or small changes in the angular momentum) or angular momentum gain (when assuming approximately fixed energy). However, the upper limit of energy also increases, and about 50\% of the streams increase their median energy. In fact, streams in the upper half of the energy distribution (larger radii) show the most gain in energy during their 6 Gyr evolution, and stellar streams at the lowest energies (with the smallest radii) show mostly small gains or small losses. Note that the rightmost panels in Figure~\ref{fig:Etot_Ltot} show the final position in energy -- angular momentum space for all particles: the particles at high energy have significantly gained, and the particles at low energy have lost or not changed their energy much. However, the maximum growth in energy is ${\sim}40000\,{\rm km}^2 {\rm s}^{-2}$ and therefore the particles with the highest final energies were initially already in the upper half of the energy distribution. Their energy gain results in a growth in their (apocenter) radii, leading to stellar streams that reach much larger radii than our initial maximum apocenter radius of $100$~kpc, which can also be seen in Figures~\ref{fig:movie} and \ref{fig:fig1} (see also Section~\ref{sec:individual}). The stream star particles that have lost energy however, originate from all throughout the initial distribution. For example, the dark blue stream that has seen a decrease that is almost as much as its initial energy.

Figure~\ref{fig:Etot_Ltot} also illustrates that while we color-code the star particles by the median change in energy for the stream they belong to, the distribution of the changes in energy can be quite large within the streams. Streams are initially clustered in the energy -- angular momentum space and spread out in both the energy and angular momentum direction, forming tendrils in this space. Moreover, the particles that have the lowest final energies have significantly lower energy than the minimum initial energy, but their colors indicate only relatively small losses, or sometimes even gains, in the median change in energy for the whole stream. The stellar stream they belong to, therefore, has to contain a significant spread in $\Delta E$.


\subsection{Origins of Stream Evolution}
To gain some understanding of the causes of the significant changes in stream morphology, orbital plane orientation, width, and energy and angular momentum, we explore their dependence on initial stream properties, but also on how close the inspiralling perturber comes to any part of a stream. We compute the closest distance between the center of mass of the perturber and any star particle of each stream at any time before the perturber has spiralled in to the central region of the host halo (after approximately 3~Gyr, see e.g. Appendix~\ref{app:orbits}).

\begin{figure}[t]
    \centering
    \includegraphics[width=0.47\textwidth]{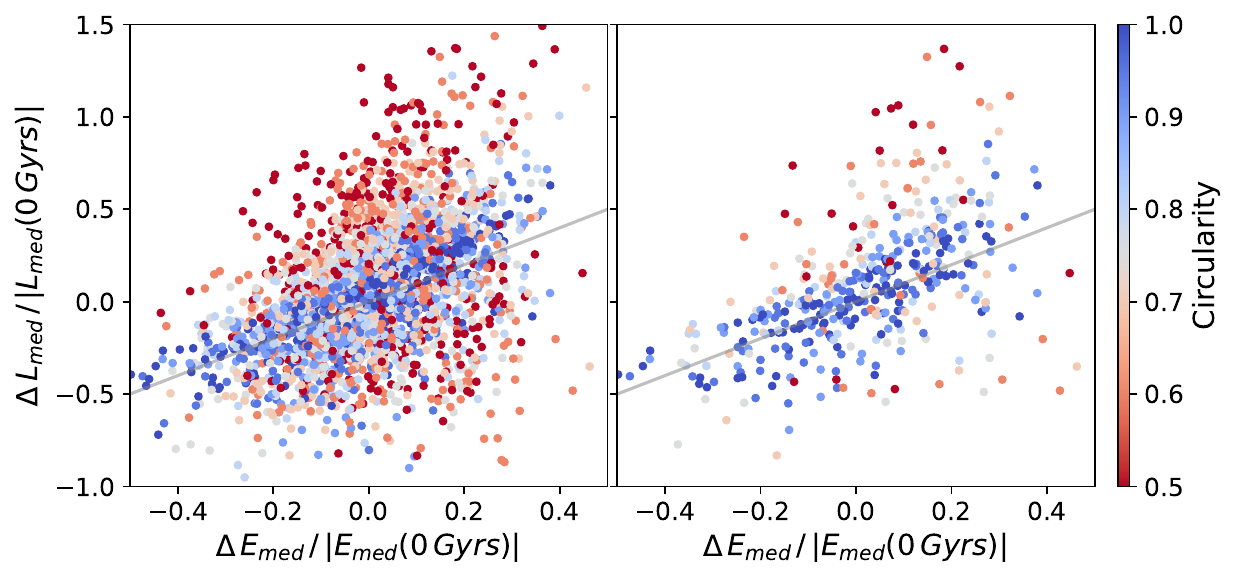}
    \caption{The relative changes in stream median energy and angular momentum, color-coded by initial orbital circularity, for all streams (left) and for streams that have \emph{no} close encounter with the perturber (i.e. the distance between stars belonging to the stream and the center-of-mass of the perturber is always $> 15$~kpc; right). Streams on more circular orbits see a correlated evolution in energy and angular momentum, while streams on more eccentric (low circularity) orbits see a significantly larger change in angular momentum, in most cases gaining relative angular momentum and thus circularizing their orbits. Note that we include streams influenced by perturbers with different orbits in this figure.}
    \label{fig:DeltaEL}
\end{figure}

Table~\ref{tab:tab3} contains the values for the closest approach and the time at which that occurred for each of the streams in our selection. Figure~\ref{fig:closest} shows for all streams the closest approach within the first 3~Gyr against the changes in stream width, orbital plane, median energy, and median angular momentum. Changes in the stream width and changes in the orbital plane correlate strongly with a close encounter with the perturber: almost all high dispersion and high angle values occur with stream -- perturber distances less than ${\sim}15$~kpc. The relative changes in energy and angular momentum however, are less correlated with the closest approach, although the most extreme values are for streams where the perturber passes closely. The points in Figure~\ref{fig:closest} are colored by the initial apocenter distance ($r_{\rm apo}$), which correlates with close stream--perturber encounters. 
This is because stellar streams with large $r_{\rm apo}$ will need to have star particles close to the infalling perturber orbit to have a close encounter, but streams with small $r_{\rm apo}$ are in a region where the perturber passes through multiple times during three pericenter passages before having sunk in. The close encounters also correlate (albeit more weakly) with circularity and stream length, which depends on stream age. This last property also correlates with apocenter radius and circularity (e.g. streams at smaller radii grow more quickly in length when measured in degrees). However, the length of the stream is also related to the initial mass of the stream: larger mass streams have a larger range in energies and can therefore become longer than low--mass streams with identical orbits and initial ages. 

\begin{figure*}[t]   
\centering
        \includegraphics[width=\textwidth]{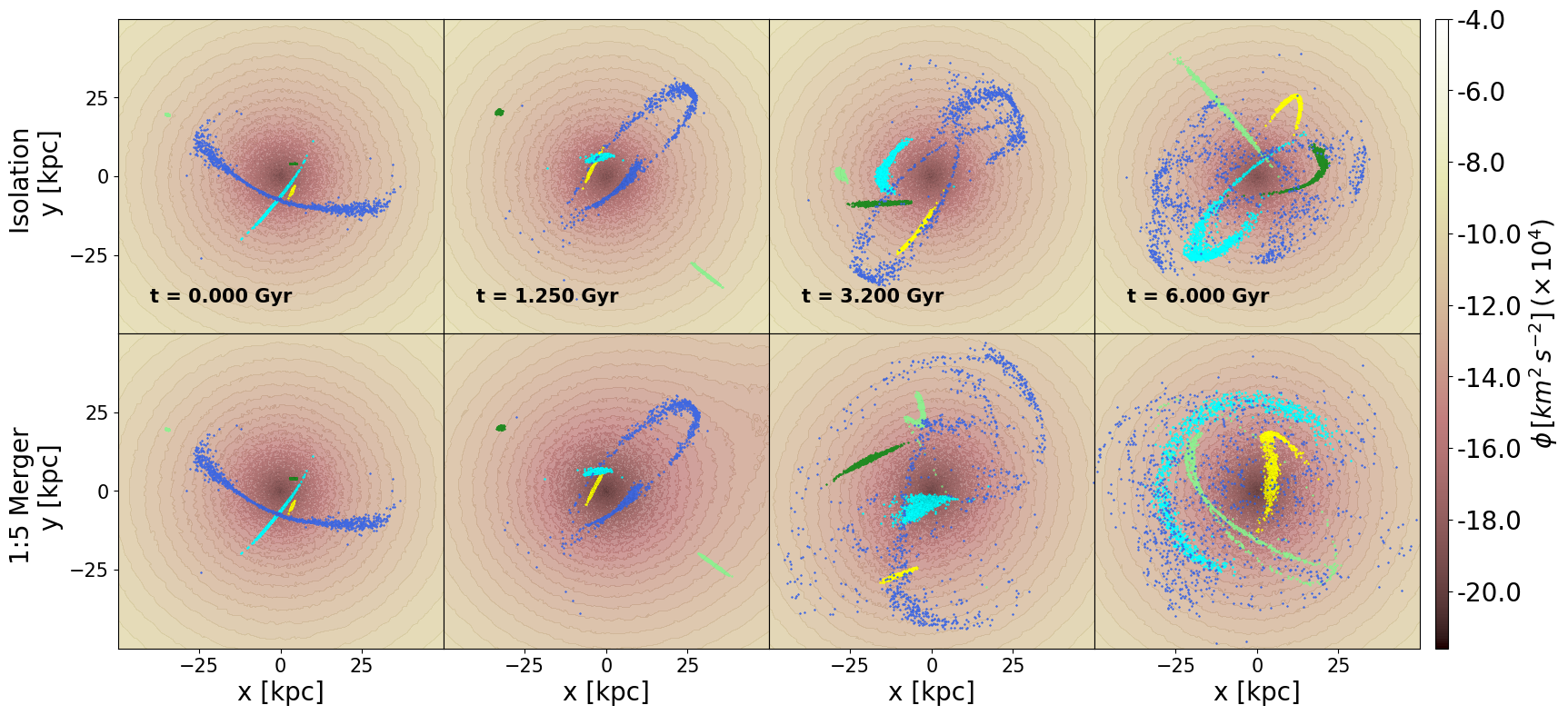}
    \caption{Evolution of 5 streams with extreme orbital evolution at (left-to-right) 0, 1.25, 3.2, and 6 Gyrs, evolved in isolation (top) and in our fiducial 1:5 merger (bottom). The background contains a contour map showing the combined potential of the host and perturber dark matter halos. Shown are streams 327 (dark green), 0 (yellow), 455 (light green), 111 (cyan), and 631 (royal blue). Significant morphological changes can be seen in all streams in the last two bottom panels, these changes start growing after 1.45 Gyrs. The dark green stream (327) is beyond 50 kpc at 6 Gyrs in the merger case.
}
    \label{fig:individual_streams_extreme}
\end{figure*}

\begin{figure*}[t]
    \centering
        \includegraphics[clip=true,trim={0cm 4cm 0cm 5cm}, width=0.95\textwidth]{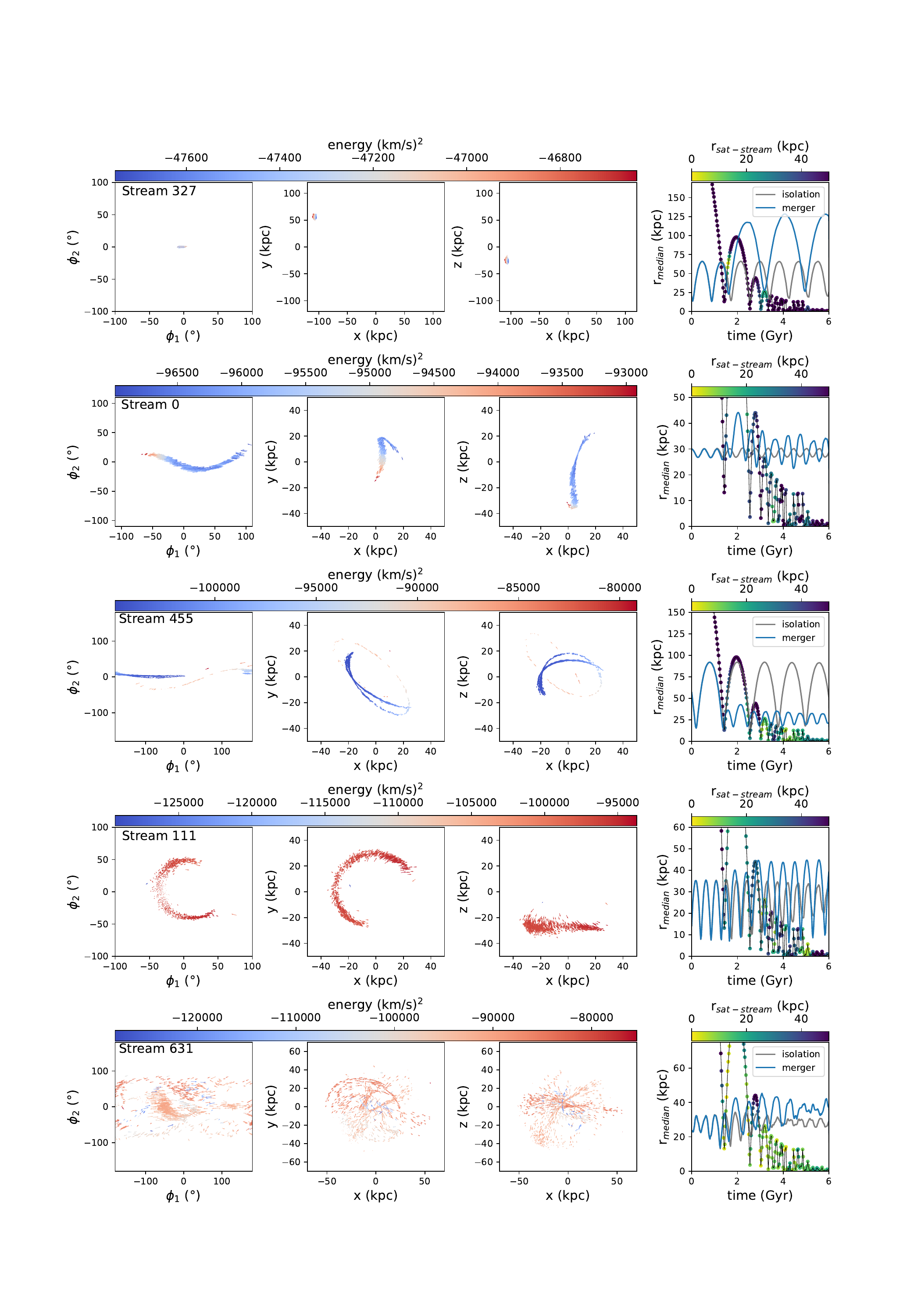}
        
    \caption{Examining streams with dramatic changes to their orbits. Positions and velocities for star particles in 5 example stellar streams in the great circle coordinate frame corresponding to the median orbit of the streams' stars (left column) and in Cartesian coordinates (x-y and x-z, middle columns) colored by total energy. The right column displays the median radius with time during the merger (blue), in the isolated halo (grey), and the radial position of the perturber (black), with the distance between the center of the perturber and the stream added highlighted in color. We note that for the stellar stream in the fourth row the transformation to the median great circle coordinates is correct. In this case all stream stars are on different radial orbits and do not appear as a cohesive stream even in the median orbital space.}
    
    \label{fig:r_timecase}
\end{figure*}

Figure~\ref{fig:closest} highlights that the stellar streams that significantly change their orbital plane or become much more dispersed are very likely to have had a close encounter with the infalling perturber. These can be different streams though than the streams that see significant changes in their energy. Taking a closer look at the changes in energy and angular momentum, we find that the circularity of orbits correlates with the change in energy and angular momentum. Figure~\ref{fig:DeltaEL} shows that the initial circularity of the stream orbit has a strong effect on the correlated change in angular momentum and energy: streams on close-to-circular orbits can gain or lose energy and angular momentum but with the relative change being approximately equal, $\Delta E_{med}/|E_{med}| \sim \Delta L_{med}/|L_{med}|$. Streams on more eccentric orbits (lower circularity) on the other hand see a larger relative gain (in most cases) or loss in angular momentum that moves away from this 1:1 relation. This effect can be seen in all streams (left panel of Figure~\ref{fig:DeltaEL}), even for streams that are never close to the perturber (right panel of Figure~\ref{fig:DeltaEL}). Streams with high circularities remain unaffected along the 1:1 relation irrespective of the distance to perturber. This shows that these streams on radial orbits are less likely to be perturbed by the merger event.

\section{The Evolution of Individual Stellar Streams}
\label{sec:individual}

After discussing how stream and orbital properties evolve during a merger, we will now describe a number of specific streams that serve as examples of particular evolutionary changes. Specifically, we look at streams with extreme orbital evolution and streams with particularly striking morphologies. We note that these two groups are not necessarily correlated: some of the streams with extreme changes in energy do not appear significantly disturbed, and some streams that show large morphological changes do not experience a large $\Delta E$.

\subsection{Streams with extreme orbital evolution}

We select five example streams that experience large changes in their orbital properties in terms of their energies and/or angular momentum. Figure~
\ref{fig:individual_streams_extreme} shows five streams with large orbital changes in the x-y plane (the orbital plane of the perturber) in isolation (top panels) and
evolved in 1:5 merger (bottom panels). Figure ~\ref{fig:r_timecase} shows the positions and morphology of these streams in both great circle and in Cartesian coordinates, each star particle color-coded by its energy and its velocity direction indicated by arrows. The fourth column shows the radial distance of the stream itself, as well as the orbit of the perturber, color-coded by the distance between the perturber and the stream.\footnote{Movies of these streams are available at \url{https://dancingstreamsinmerginghalos.github.io}. Detailed simulation data on one or more stellar streams are available on reasonable request.} Most of these streams see relatively abrupt changes in their orbits following close encounters with the perturber, either during its first pericenter passage, or during its first and second pericenter passages. Below, we briefly describe the evolution of these, and list their initial properties.

\begin{figure*}[t]   
\centering
        \includegraphics[clip=true,trim={4cm 1cm 7cm 3cm },width=\textwidth]{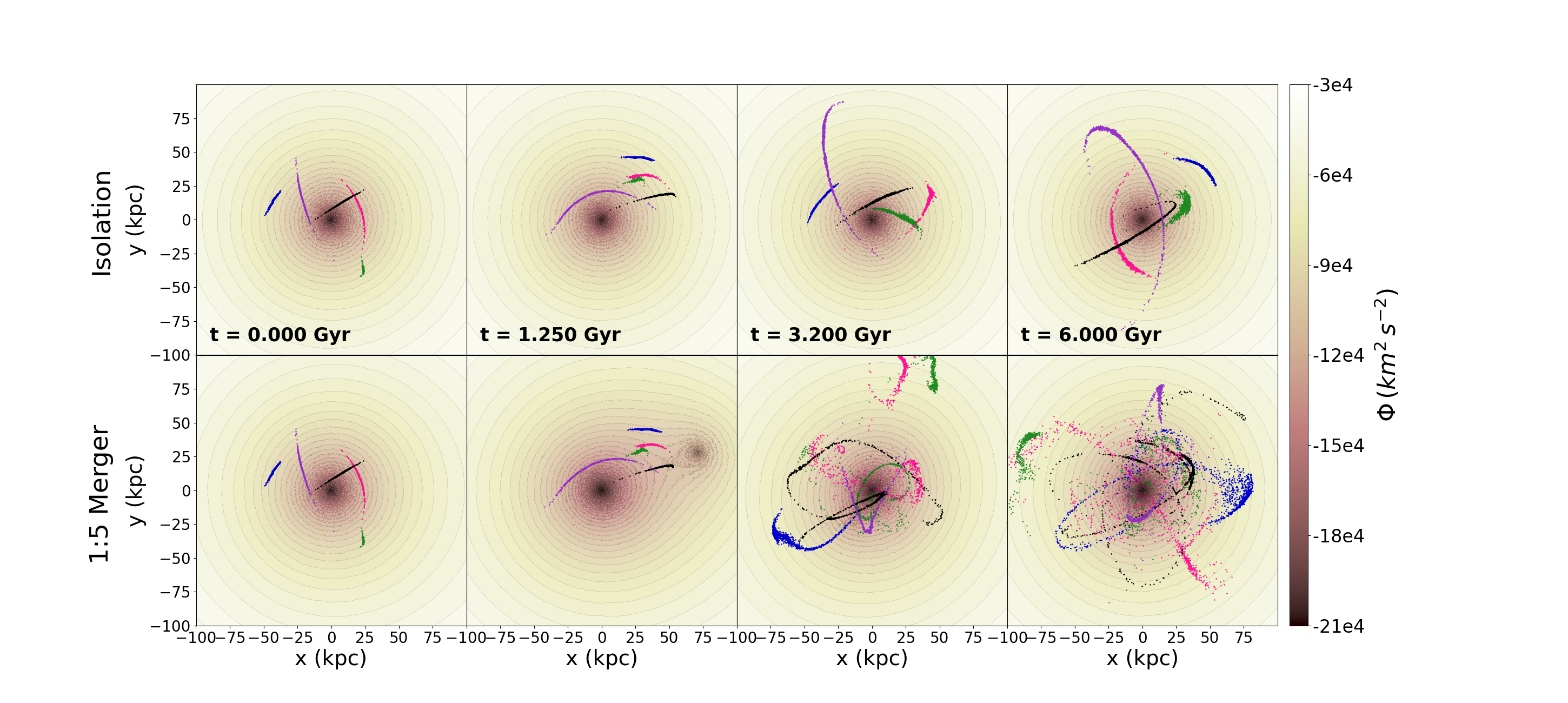}
    \caption{Evolution of 5 streams with striking morphological behavior at (left-to-right) 0, 1.25, 3.2, and 6 Gyrs, evolved in isolation (top) and the streams in 1:5 merger (bottom). The background is a contour map showing the combined potential of the host and perturber dark matter halos. Shown are streams 175 (green), 232 (dark blue), 243 (pink), 444 (black), and 502 (purple). Striking morphological changes to streams are seen after 1.45 Gyrs: the pink (in the third panels) and green (in the fourth panels) streams separate in two components, the purple (fourth column) and black (third column) stream develop sharp turnarounds, the purple and blue stream appear to radialize, the black appears to develop bifurcations (fourth column) and the blue stream develops an off-centered shell (fourth column).
}
    \label{fig:individual_streams}
\end{figure*}

\begin{figure*}[t]
    \centering
        \includegraphics[clip=true,trim={0cm 4cm 0cm 5cm}, width=0.95\textwidth]{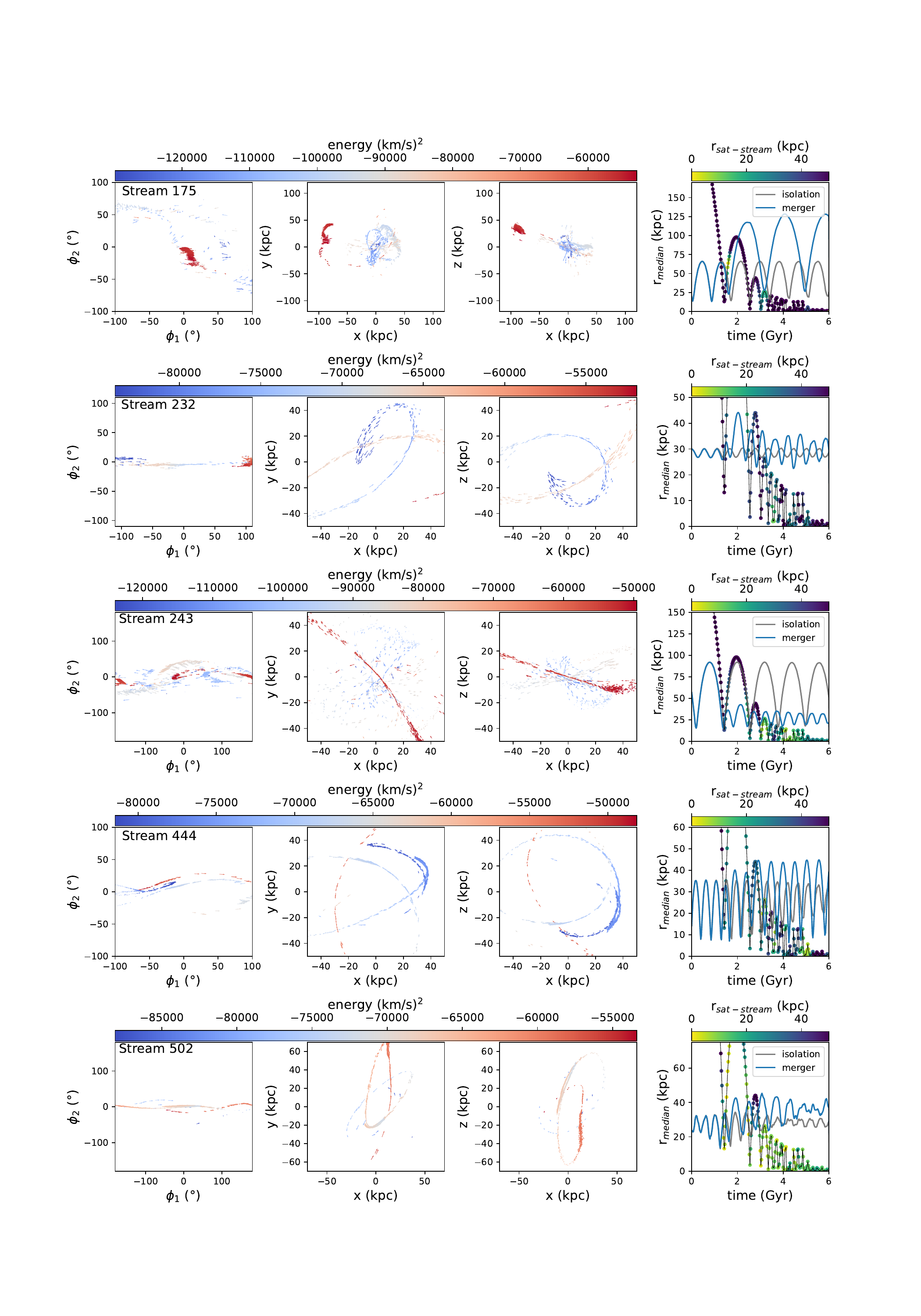}
    \caption{Final positions and velocities for star particles in the 5 example stellar streams for which the evolution is shown in Figure~\ref{fig:individual_streams}: great circle coordinates (left column), Cartesian coordinates (x-y and x-z, middle columns colored by total energy, and their median radius with time (right column) during the merger (blue) and in the isolated halo (grey) and the perturber radius (black) with the distance between the perturber and the stream added (colored points).
}
    \label{fig:Fig9streams_rtime}
\end{figure*}

\begin{enumerate}
    \item [Stream 327] (dark green in Figure~\ref{fig:individual_streams_extreme}; age $= 0.5$~Gyr, $r_{\rm apo} = 80$~kpc, $\eta = 0.5$) This short, low--mass stream on a fairly radial orbit has the largest increase in median energy over the 6 Gyr merger simulation, relative to its initial energy. After a close encounter with the perturber around the perturber's first pericenter passage and just after the stream's pericenter, this stream gets dragged outward by the perturber and reaches a substantially larger orbit with an apocenter radius of ${\gtrsim}120$~kpc. 
    \item [Stream 0] (yellow; age $=$0.5~Gyr, $r_{\rm apo} = 30$~kpc, $\eta = 1.0$). This short, circular, low--mass stream with a small initial apocenter moves to larger radii when both stream and perturber move outward after pericenter passages, with a subsequent drop back to a similar apocenter radius as that which it started with. However, that only describes the median trend of the stream's evolution. Even though the stream is young and quite short initially, the range in energy and angular momentum of the stream particles increases substantially and there arises a gradient with the trailing arm on gradually more and more radial orbits, and the leading arm on less radial orbits. The trailing arm overtakes the leading arm and the stream becomes a combination of stream- and shell-like aspects.
    \item [Stream 455] (light green; age $=$0.5~Gyr, $r_{\rm apo} = 100$~kpc, $\eta = 0.5$).  A radial, low--mass stream with an initial apocenter of 100 kpc which loses the most energy over 6 Gyrs. We note that in contrast to Stream 327, here the stream particle precedes the perturber in pericenter and therefore may have encountered a significant additional gravitational pull from the central region just after having passed through pericenter. This may have led to decreasing the stream's energy and shrinking but circularizing its orbit. 
     \item [Stream 111] (cyan; age $=$~4 Gyr, $r_{\rm apo} = $~40 kpc, $\eta = $0.5). This somewhat radial stream with a small apocenter fully evolves into a shell during the merger. The star particles along the stream get pulled into more and more radial orbits by pericenter passages of the perturber of which the first one is when the stream is at apocenter and at its most extended. At the final snapshot all star particles have close to fully radial orbits but, as they started the apocenters of their radial orbits at different phases along their original orbit, the stars are not on the same orbital plane, and form large shells in both physical and orbital space.
     \item[Stream 631] (royal blue; age $=$~5 Gyr, $r_{\rm apo} = $~40 kpc, $\eta = $0.5). Having a close-to-circular, close-in orbit initially, this stream wraps around but then gets buckled by the perturber. The different stream wraps end up on different orbits and disperse, filling a large part of the median orbital plane as well as the cartesian space at radii smaller than the original $r_{\rm apo}$.
\end{enumerate}
These individual cases suggest that if the perturber and stream have a small offset in orbital phase the changes in orbit may be largest. There is however, no correlation between sign($v_r$) of the perturber and of the stream. Some of these streams show large variations in morphology while others do not.

\subsection{Morphologically Striking Streams}
\label{sec:morphology}
Following streams with significant orbital changes, we now explore some of the streams that develop notable morphological features.

Figure~\ref{fig:individual_streams} shows five streams in the x-y plane (the orbital plane of the perturber) in isolation (top panels) and evolved in 1:5 merger (bottom panels). Figure~\ref{fig:Fig9streams_rtime} shows these same streams in more detail, including their great circle coordinates (left column) and the stream median orbits and distance from the perturber (right column). Below, we shortly describe the evolution of these streams.
\begin{enumerate}
    \item [Stream 175] (green in Figure~\ref{fig:individual_streams}; age $=$~4~Gyr, $r_{\rm apo} = 50$~kpc, $\eta = 0.5$) In the merger case, this stream splits around 3.2 Gyrs, after which the two parts evolve on different orbits. One part follows a somewhat similar orbit to the stream in isolation, while the second part of this stream moves to a more radial orbit with a larger apocenter ($\gtrsim 100$~kpc). The distribution of energy in the stream also diverges following the morphological split.
    \item [Stream 232] (dark blue; age $=$~4~Gyr, $r_{\rm apo} = 60$~kpc, $\eta = 1$) This stream with a circular orbit in isolation moves to a much less circular orbit with the apocenter radius increasing to ${\sim}90$~kpc after 1.5 Gyrs. Around 3.2 Gyrs the stream has a shape consisting of two arcs with one denser and developing into a shell-like structure.
    \item [Stream 243] (pink; age $= 5$~Gyr, $r_{\rm apo} = 60$~kpc, $\eta = 0.8$) This stream evolves into a narrow, long stream in isolation, but in the 1:5 merger one part of the stream is expelled outward around 3.2 Gyrs. The apocenter radii of stars in this part of the stream change from ${\sim}50$ kpc to $100$--$130$ kpc.
    \item [Stream 444] (black; age $= 6$~Gyr, $r_{\rm apo} = 90$~kpc, $\eta = 0.75$) The perturber interacts with the trailing arm of this stream, pushing the trailing arm outward, but creating folds in the leading arm and pulling it inward. This leads to a large gradient in orbits for the stars in the stream, where the leading arms appear to orbit with a smaller apocenter than initially, while the trailing arms have a larger apocenter radius than the initial stream. 
    \item [Stream 502] (purple; age $= 5$~Gyr, $r_{\rm apo} = 100$~kpc, $\eta = 0.6$) This initially thin and long stream develops kinks and folds during the merger and moves to a more circular orbit with a somewhat smaller apocenter. Because of this, the stream grows significantly more in length in the merger case where it wraps around at least once while it does not wrap around yet in the isolated simulation.
\end{enumerate}  

These example cases demonstrate that individual streams undergo significant morphological changes during a major merger. Streams may appear completely disconnected to an observer, despite sharing the same origin.

\section{Discussion}\label{sec:discussion}

Several previous work show that interactions with subhalos \citep[e.g.][]{Bonaca+2019,Carlberg+2020}, accretion of streams from a dwarf galaxy onto a MW--like host \citep[][]{Malhan19, Qian+22}, disk tilting \citep{Nibaur+2024}, orbits on dynamical separatrices \cite{Yavetz+2021}, and the presence of large subhalos like Sagittarius \citep{Dillamore+2022} or the LMC \citep{Brooks+2024} in the MW halo can affect the appearance of streams, for example producing gaps or under densities, bifurcations and/or other substructures. 
These theoretical results are supported by observational findings of new or known stellar streams with detected substructures, reflecting their complex nature \citep[e.g.][]{Bonaca2019,Ferguson2022, Nidever2023}.

Within this context, this study takes a step back and considers the overarching question of how streams evolve under halo (galaxy) mergers. In our merging system, we see streams exhibiting similar substructures as identified before, such as misalignments, gaps, bifurcations or otherwise splitting into fragments, and (partially) dissolvement. However, we show that these morphological indicators can also coincide with changes in stream orbits, both with respect to the circularity of the orbit and its radial extend and to the orbital plane. In addition, we see streams that show large changes in their orbit but no morphological substructure to indicate their disturbed history.

As described in Section~\ref{method} all our stellar streams have a fixed number of particles, and no new stream particles are generated during the merger simulation. While this leads to underdensities in the central region of the stellar streams, this does not appear to contribute significantly to the splitting and bifurcations of streams that we observe: in all of those cases parts of a stellar stream move onto different orbits, including different orbital planes, and different apocenter and pericenter radii. Moreover, the same limitation applies to our isolated halo simulations but no gaps appear in the stellar streams evolved in those simulations.

\subsection{Future directions}
While simulating an LMC--MW interaction to study stream evolution allows comparisons to existing theoretical and observational results, the strong effects we see suggest the question of  how streams in different mergers would evolve. In Weerasooriya et al. (in prep.) we compare the stream properties in different mass ratio mergers.
Additionally, it is important to note that our simulation does not account for the effects of other satellites orbiting the MW or being accreted, or for the actual baryonic components of the galaxy itself. The presence of additional satellites could further influence the evolution of stellar streams as detailed by \cite{Arora+23}. Moreover, \citet{Nibaur+2024} have shown that a tilted disk (possibly tilted due to a previous merger) can also generate fanning and narrowing of streams. While we do not have a disk potential in our simulations, some of the deformations of the dark matter halo could create similar gravitational effects. Future studies should explore how stellar streams evolve in a more complex environment with evolving galaxy potentials and multiple satellites. 

\section{Summary \& Conclusions}\label{sec:conclusions}

In this work, we have explored the impact of major mergers on stellar stream properties after the perturber has completely merged in a MW--LMC-like merger. We evolved 1024 stellar streams with known systematically varied initial conditions for 6 Gyrs in a MW potential in isolation and in the presence of an infalling perturber using high-resolution $\mathrm{N}$-body simulations.

\subsection{Summary of Findings}

\begin{itemize}
    \item We find that streams with the same initial conditions change significantly both morphologically and in energy space in the presence of a MW--LMC-like merger in comparison to the same streams evolved in isolation. 
    
    \item Streams in the inner halo (within 50 kpc) get significantly dispersed in comparison to those in the outer halo due to the profound steepening of the potential gradient during the merger.
    
    \item Stellar streams show distinctive sub-structures as a result of the interaction with the perturber. Higher--mass stellar streams are more likely to have  oscillatory substructures (``waves") in the great circle coordinate frame.

    \item The location of the orbital pole, or orientation of the orbital plane, can change significantly for streams that have evolved through a merger. As a result, their position in the sky experience a notable shift. This re-orientation does not always coincide with notable morphological changes.
    
    \item Some stellar streams undergo extreme orbital/energy changes, and those that do can change e.g. their orbital circularity or median energy in either positive or negative direction for energy or circularity. While some of these also have significant apparent deformations in their morphology, others do not.
    \item Stellar streams can develop feather and bifurcations or split into multiple parts that may have different orbits. This suggests that observed streams that appear disconnected at present day may have the same origin. Interpretations of stream width and stream length in relation to a progenitor mass and a stream age may thus be misleading.

    \item Close approaches to the perturber correlate strongly with the change in the orbital plane and the increase in stream width and show some correlation with relative changes in energy and angular momentum.

    \item The relative changes in energy and angular momentum are strongly correlated for streams that have close-to-circular orbits initially, whether or not they experience a close encounter with the perturber. Streams on radial orbits that experience a close encounter with the perturber do not follow this correlation. These can gain significant angular momentum and thus move toward more circular orbits.
\end{itemize}

Given the LMC is currently perturbing the MW, even though in quite early stages of its merger, some of the stream evolution identified in this work may already be relevant for interpreting the origin of streams in the MW that are likely to have been perturbed by LMC. For example, we highlight the possibility that some stellar streams in the MW, identified by observers as distinct, may share a common origin. These streams may have originally evolved as a single entity but later diverged due to interactions with the perturber. The Orphan Chenab (OC) streams are a great example of  streams that were discovered separately but are now thought to be two parts of one stream \citep{Grillmair2006,Belokurov+2006,Shipp+2018}. This work highlights that there may well be additional cases where the streams' present-day orbits deviate more strongly.

\subsection{Overall Conclusion}
Our findings highlight the significant impact of a major accretion event on the morphology, energy distribution, and orbital evolution of stellar streams, emphasizing that interactions with a perturber can lead to complex changes in stream structure, kinematics, orbit and sky position. These results emphasize limitations on the use of streams to infer the properties of their host galaxy's potential and history without accounting for their evolution in a hierarchical context. At the same time, they emphasize how each stream provides a unique perspective on the (at times) stochastic evolution of the host. Understanding a galaxy's {\it system} of streams, rather than modeling individual streams, therefore can enable surpassing the limitations implied by this study. Moreover, while we refer to the merger discribed in this work as an MW--LMC-like interaction, our results are perhaps even more relevant for interpretations of stellar streams in external galaxies that may have experienced past major mergers. A detailed study of stellar stream evolution in different mass ratio merger will be described in future work.

\begin{acknowledgements}
The authors thank anonymous reviewers for their valuable suggestions and help in strengthening this manuscript. This project started during the ``Big Apple Dynamics'' 2021 Summer School on galactic dynamics at the Center for Computational Astrophysics of the Flatiron Institute. The simulations were run using \textsc{RUSTY} computer cluster at CCA, Flatiron Institute, NY. The Flatiron Institute is supported by the Simons Foundation. We would like to thank Jason Hunt for hosting our simulation data. SW would like to thank Helmer Koppelman for his guidance in running the simulations during Galactic Dynamic Summer School 2021.
KVJ was supported by Simons Foundation grant 1018465. TS gratefully acknowledges support by NASA grants 22-ROMAN22-0055 and 22-ROMAN22-0013, NSF grant AST-2421845 and Simons Foundation grant MPS-AI-00010513.
\end{acknowledgements}

\bibliography{streams}{}
\bibliographystyle{aasjournal}

\appendix
\label{Appendix}
\section{Impact of Resolution}
\subsection{Testing resolution and numerical heating: Stellar Streams in Isolation}
\label{app:heating}
In this section we describe in more detail how stellar streams are impacted when evolved in an isolated MW--like host halo for 6 Gyrs. This allows us to understand the effects of stellar stream properties due to numerical heating during their evolution and other effects in the absence of a major merger.

\begin{figure}[!ht]
    \centering
    \includegraphics[clip=true,trim={1cm 0cm 1cm 0cm},width=0.45\textwidth]{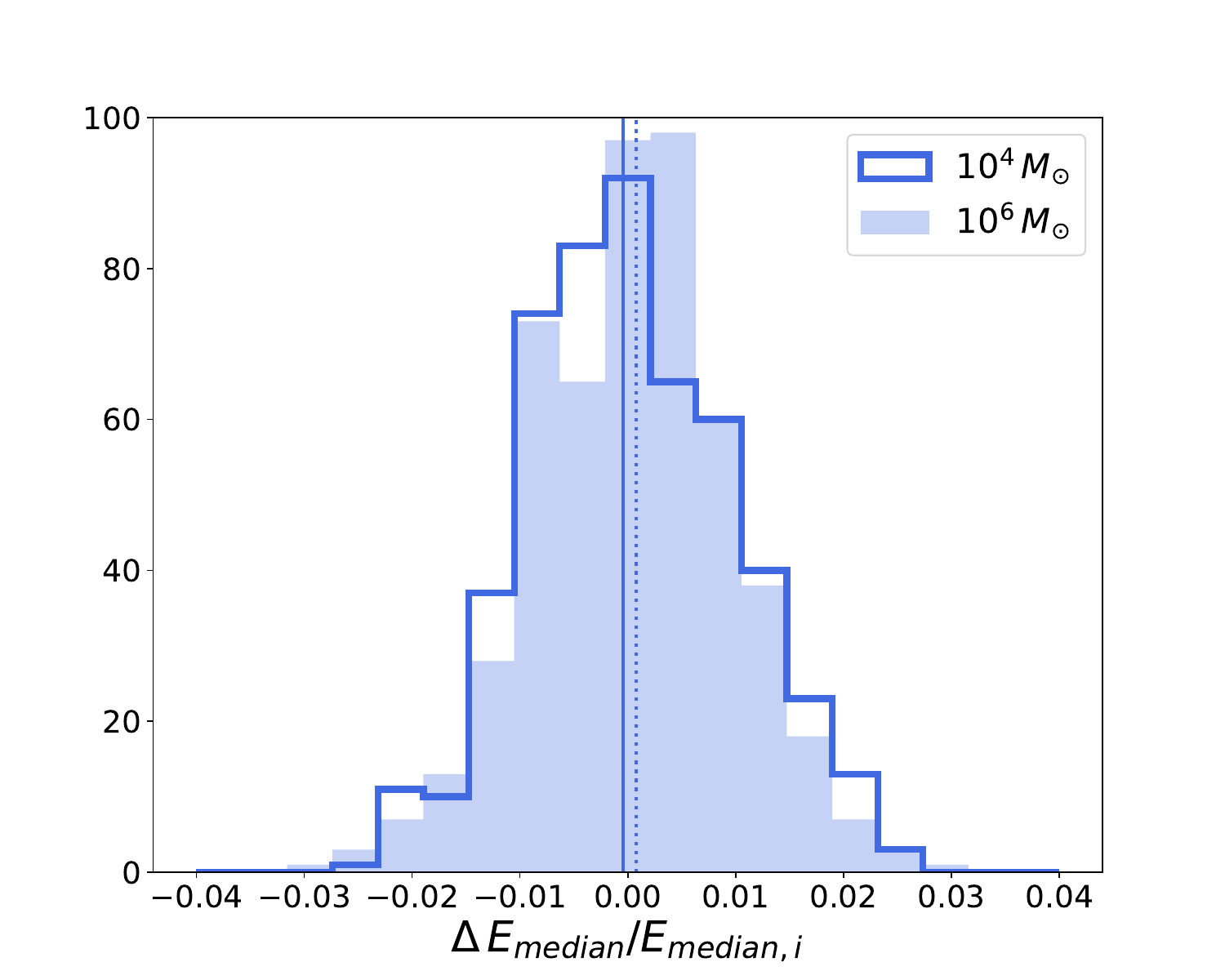}
\hfill
\includegraphics[clip=true,trim={1cm 0 1cm 0},width=0.45\textwidth]{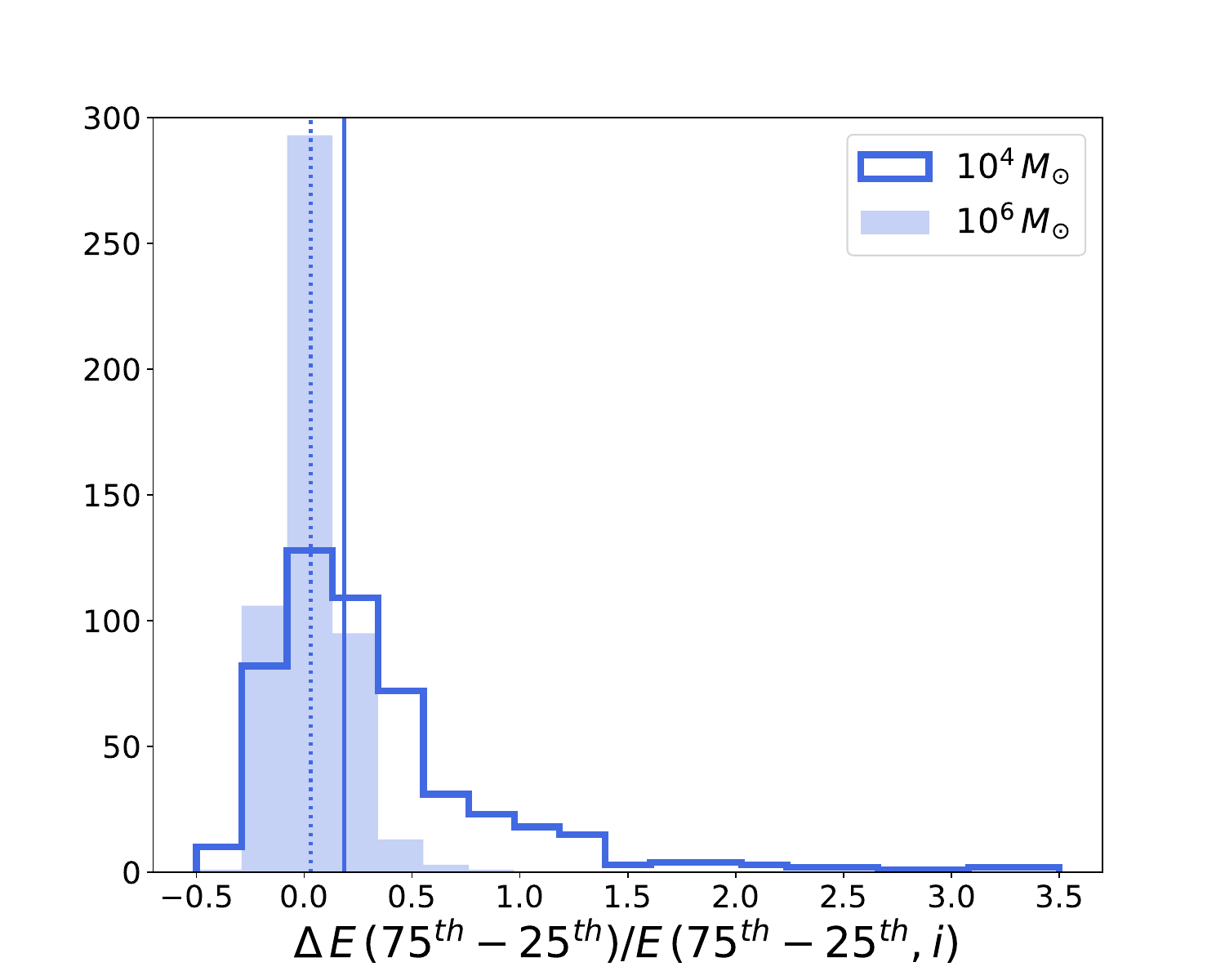}
    \caption{The distribution of changes in the total energy of star particles in stellar streams. Left panel: the relative distribution of the change in median energy per stellar stream compared to the initial values. Right panel: the distribution of the change in the interquartile range (IQR) of the total energy for stars within each stellar stream, with respect to the initial IQR. Separate distributions are shown for low--mass streams (solid line) and high--mass streams (shaded area).}
    \label{fig:1haloEmed}
\end{figure}
Figure~\ref{fig:1haloEmed} shows the change in the distributions of the median and interquartile ranges (IQR) for the energy of star particles in each stellar stream between the the beginning and end of a 6~Gyr evolution in isolation. The median energy of star particles in stellar streams barely changes ($<3\%$) over 6~Gyr of evolution in a live N-body halo for both low--mass and high--mass streams. However, the distributions of energy IQR do change somewhat with a tail towards growing IQR for the energy of star particles in stellar streams, most notably for low--mass streams (up to more than doubling the IQR for a small fraction of streams).  This indicates that we see some minimal heating of the low--mass stellar streams. However, this numerical heating is negligible in comparison to the changes in the energy distributions seen in merger simulations as shown in Figure~\ref{fig:Emed}.

\begin{figure}[!ht]
    \centering
    \includegraphics[width=\linewidth]{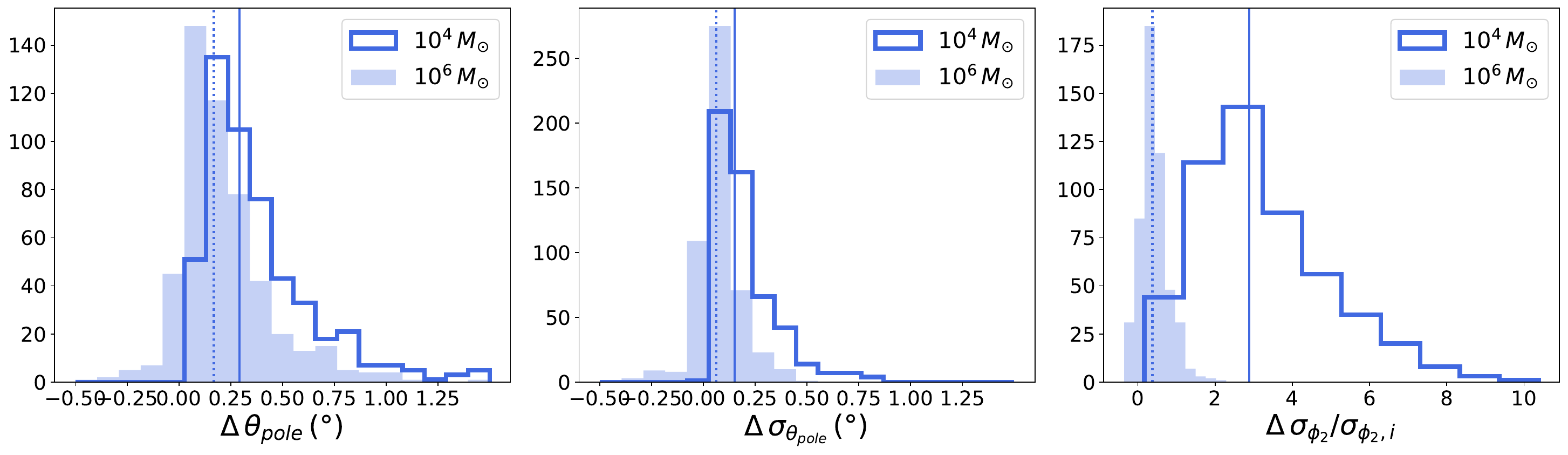}
    \caption{Left panel: distribution of change in orbital pole for stellar streams evolved in isolation. Mid panel: Distribution of change in dispersion of the orbital pole. Right panel: distribution of relative change in thickness ($\phi_2$) in GC coordinates for streams evolved in isolation. Low--mass streams are shown in the line histogram while the high--mass streams are shown by the shaded regions.}
    \label{fig:distributions}
\end{figure}

Additionally, Figure \ref{fig:distributions} shows the changes in pole angles, the growth in the standard deviation of the distribution of pole angles for individual star particles in stellar streams, and the relative change in thickness between 0 and 6 Gyrs shown by left, middle, and right panels, respectively. All of these measures show some effects of numerical heating for low--mass streams similar to Figure~\ref{fig:1haloEmed}. The low--mass streams show a slightly larger dispersion in orbital plane within the streams but grow by a factor of a few in thickness compared to high--mass streams when evolved in isolation. This is likely due to the initial thinness of the low--mass stellar streams, making any, no matter how small, effect of numerical heating noticeable in the relative thickness and energy range of the streams. Additionally, because of their larger initial energy ranges, the higher-mass streams will experience more elongation, which can result in a relative thinning. This last effect may also be slightly exaggerated since our streams do not gain star particles during their evolution.
We emphasize that even though we see small effects due to numerical heating for the low--mass streams, these stand in no comparison to the changes we see in the merger simulation as described in Section~\ref{sec:15stat}.

\subsection{Testing resolution and numerical heating: Stellar stream structure}
\label{app:heating1}

\begin{figure*}[h!]
\centering
    \includegraphics[width=.7\textwidth]{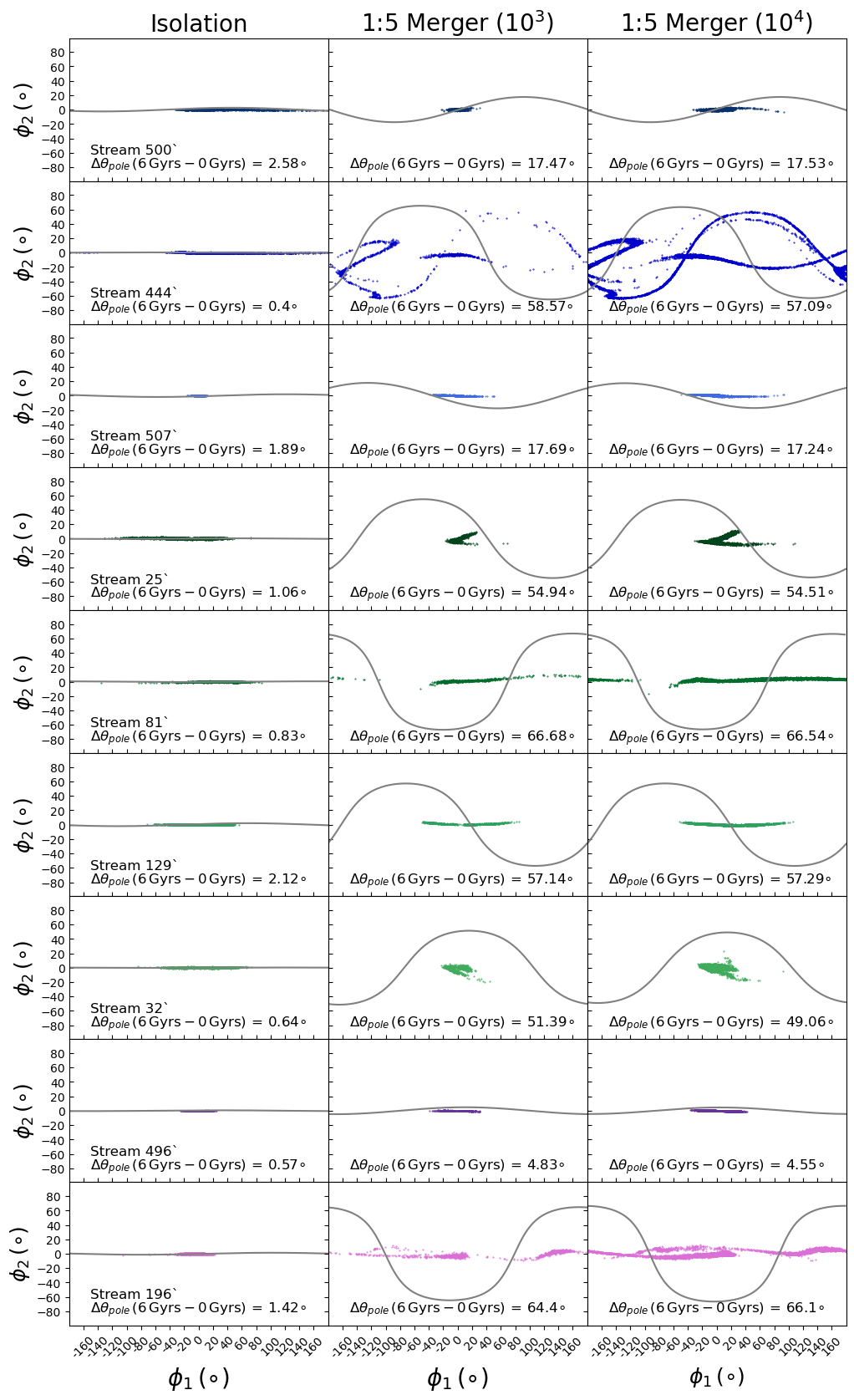}
    \caption{The final stream positions, oriented according to the stream’s orbit, for nine $10^4\,\mathrm{M_{\odot}}$ streams with ages, apocenter radii, and orbital circularities identical to the streams shown in Figure \ref{fig:fig3}. Each stream is colored by their group similar to Figure \ref{fig:fig3}. Each row shows one stellar stream, evolved in isolation (left column), evolved with 1000 particles in the stream during a 1:5 merger (middle column), and evolved with 10,000 particles in the stream during a 1:5 merger (right column). Streams evolved in a 1:5 merger form oscillatory patterns similar to Figure \ref{fig:fig3}. The higher resolution stream more densely samples the stream spatial distribution and shows these substructures more clearly.}
    \label{fig:higherstreamres1}
\end{figure*}

\begin{figure*}[h!]
\centering
    \includegraphics[width=.75\textwidth]{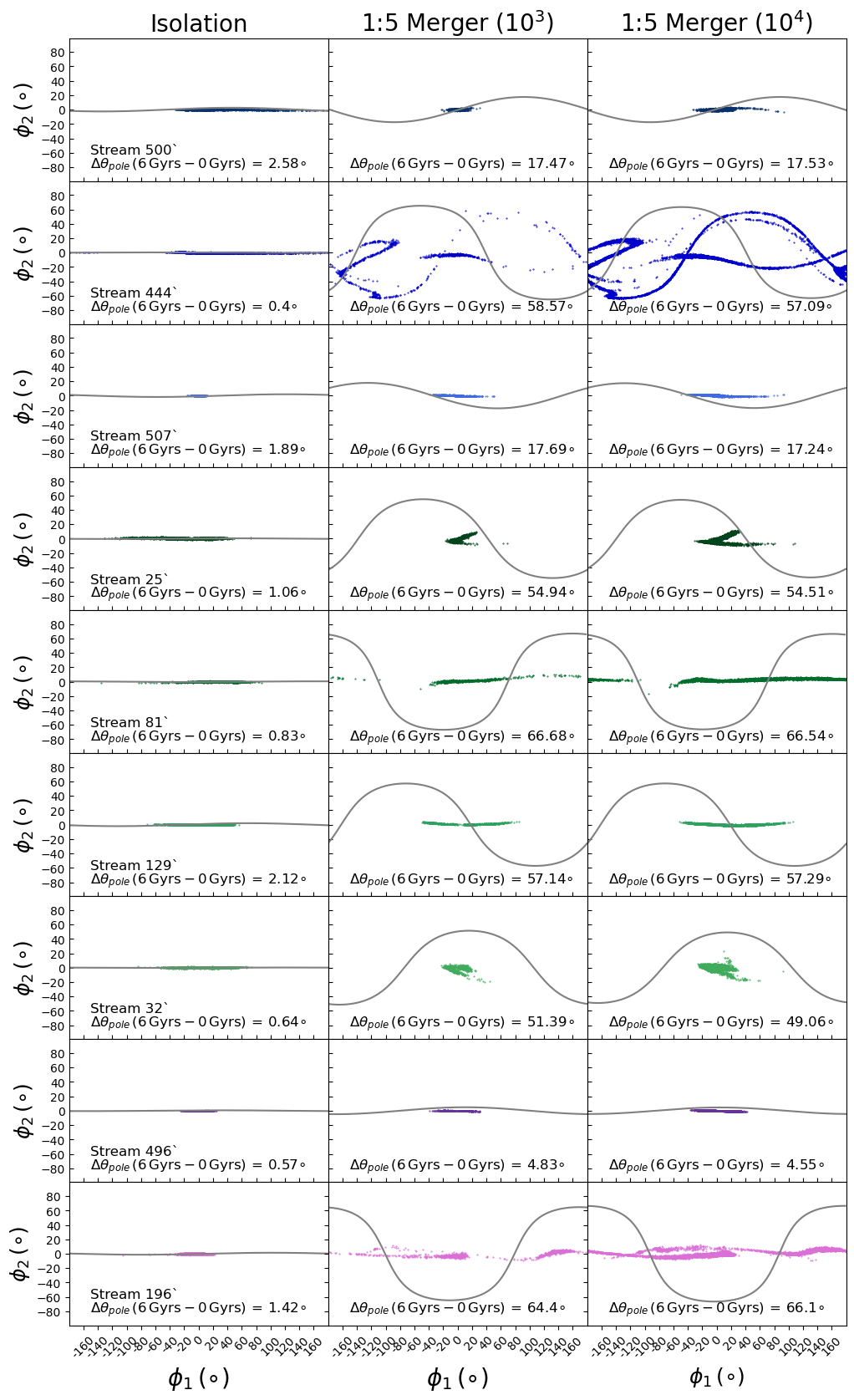}
    \caption{The final stream positions, oriented according to the stream’s orbit, for nine $10^6\,\mathrm{M_{\odot}}$ streams with ages, apocenter radii, and orbital circularities identical to the streams shown in Figure \ref{fig:fig3}. Each stream is colored by their group similar to Figure \ref{fig:fig3}. Each row shows one stellar stream, evolved in isolation (left column), evolved with 1000 particles in the stream during a 1:5 merger (middle column), and evolved with 10,000 particles in the stream during a 1:5 merger (right column). Streams evolved in a 1:5 merger form oscillatory patterns similar to Figure \ref{fig:fig3}. The higher resolution stream more densely samples the stream spatial distribution and shows these substructures more clearly.}
    \label{fig:higherstreamres2}
\end{figure*}

In order to test the robustness of the stellar stream structure and evolution to resolution within the streams, we have completed additional 1:5 merger simulations, identical to the one described in this paper. For these additional simulations we focus on 18 stellar streams (nine $10^4\,\mathrm{M_{\odot}}$ and nine $10^6\,\mathrm{M_{\odot}}$ streams) with ages, apocenter radii, eccentricities, and mass set to the same values as for the streams shown in Figure~\ref{fig:fig3}, either at the same resolution as in our fiducial merger simulation (1,000 particles in each stream) or with ten times higher resolution (10,000 particles in each stream). We find that the evolution of the streams and any characteristics that we describe in the main text to be robust. Figures~\ref{fig:higherstreamres1} and ~\ref{fig:higherstreamres2} show that as discussed in Section~\ref{selection}, while all streams form oscillatory patterns, higher mass streams seem more likely to form oscillatory substructures. Higher mass streams are also more disturbed and thicker in comparison to low mass streams. A higher particle resolution in the streams allows the star particles to more completely occupy the streams' phase space distribution, but the particles still follow the same tracks as the low particle resolution simulation. This validates our results and provides further support for the conclusions described in the main text.

\section{Perturber Orbits}
\label{app:orbits}

\begin{figure}[h!]
    \centering
    \includegraphics[width=10cm]{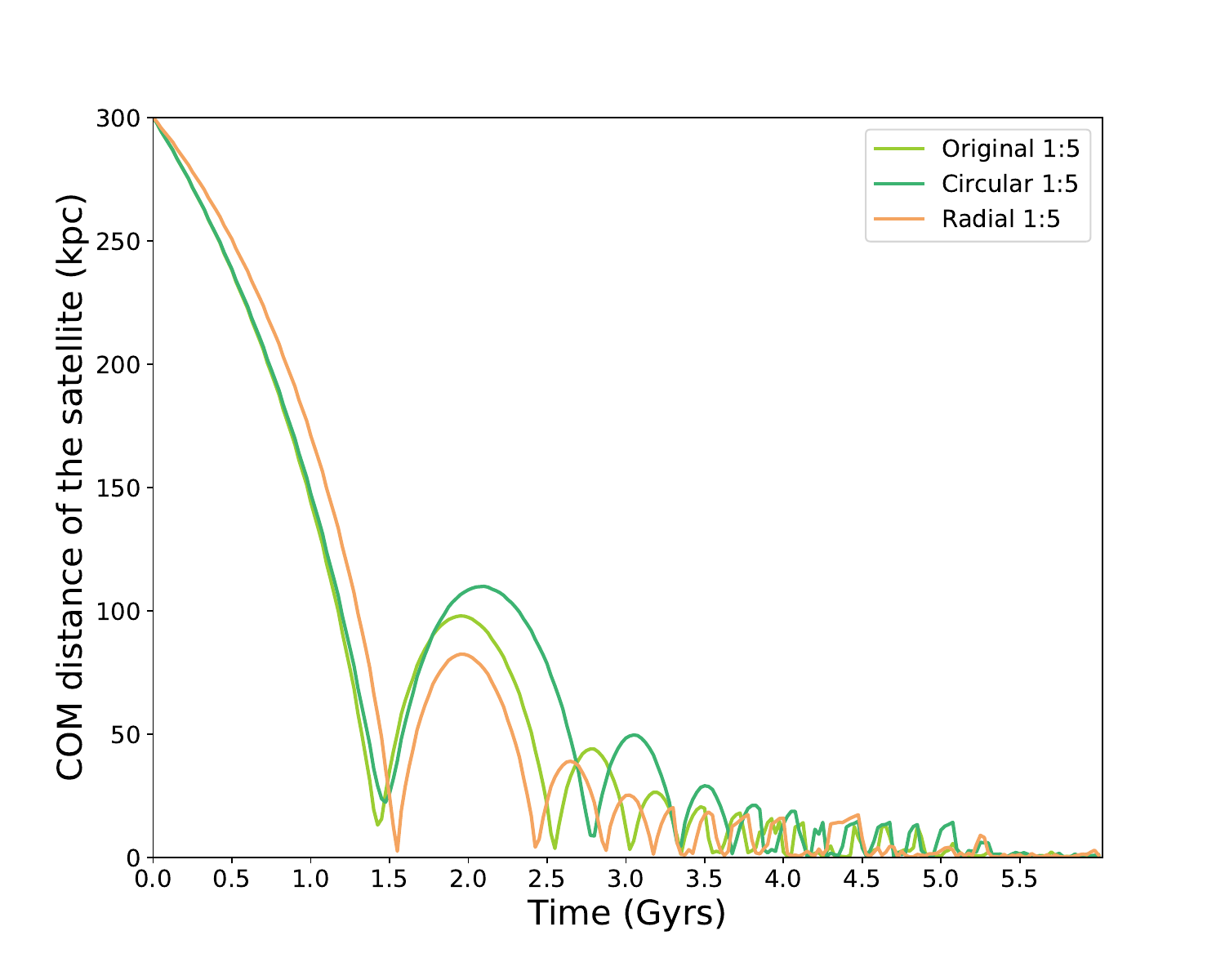}
    \caption{The orbital evolution of the (identical) perturber on three diferent orbits in a 1:5 merger: the original orbit (light green), a more circular orbit (green), and a more radial orbit (sandy brown).}
    \label{fig:r_time_satelliteorbits}
\end{figure}

Figure \ref{fig:r_time_satelliteorbits} shows the orbital evolution of identical perturbers on different orbits: the original (fiducial) orbit, a more circular orbit, and a more radial orbit. The perturber initial position is always set to $r_{sat,0}=[300,0,0]\, \text{kpc}$, and the initial velocities are set to: $v_{sat,0}=[-100,30,0]\,\text{km} \text{s}^{-1}$ (fiducial; $\eta=0.27$), $v_{sat,0}=[-100,45,0]\,\text{km} \text{s}^{-1}$ (circular; $\eta=0.39$), and $v_{sat,0}=[-75,0,0]\,\text{km} \text{s}^{-1}$ (radial; $\eta=0.0$). Even though the radial infalling velocity is initially lower for the perturber on the radial orbit, as expected the perturber with a more radial orbit merges faster than the original orbit, which merges faster than the circular orbit.

\begin{figure}[!ht]
    \centering
    \includegraphics[clip=true,trim={0cm 0cm 0cm 0cm},width=\textwidth]{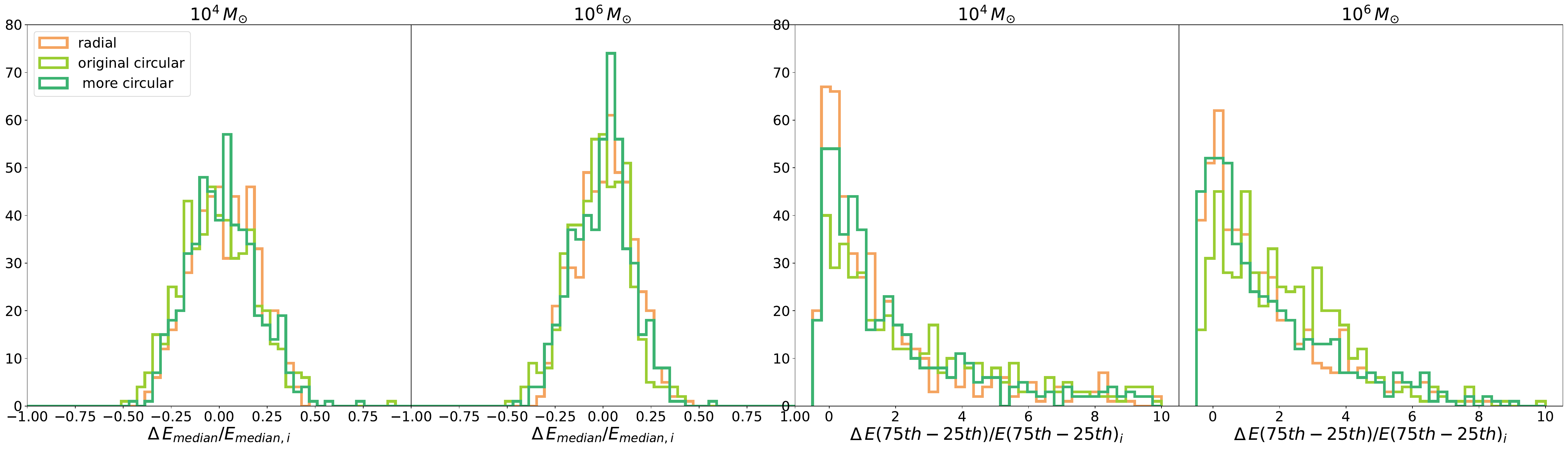}
    \caption{The ratio of distribution of median energy change with respect to initial median energy at 0 Gyrs (two leftmost panels) and ratio of change in inter-quartile range w.r.t. IQR in energy at 0 Gyrs (two rightmost panels) relative to their initial values, for low--mass (first and third panel) and high--mass (second and fourth panel) stellar streams evolved in our three 1:5 merger simulations.}
    \label{fig:Emed}
\end{figure}

Figure \ref{fig:Emed} shows the distribution of change in median energy between 0 and 6 Gyrs for low and high--mass streams (1st two plots from left to right), and the distribution of change in the interquartile range of stellar streams (3rd and 4th panels) for all three perturber orbits. Overall, the distributions of these quantities are minimally affected by the change in orbit of the perturber, but there may be somewhat more streams that see little change in the interquartile ranges of their energy distributions for the radial orbit.

\begin{figure*}
    \centering
        \includegraphics[clip=true,trim={0cm 4cm 0cm 5cm}, width=0.95\textwidth]{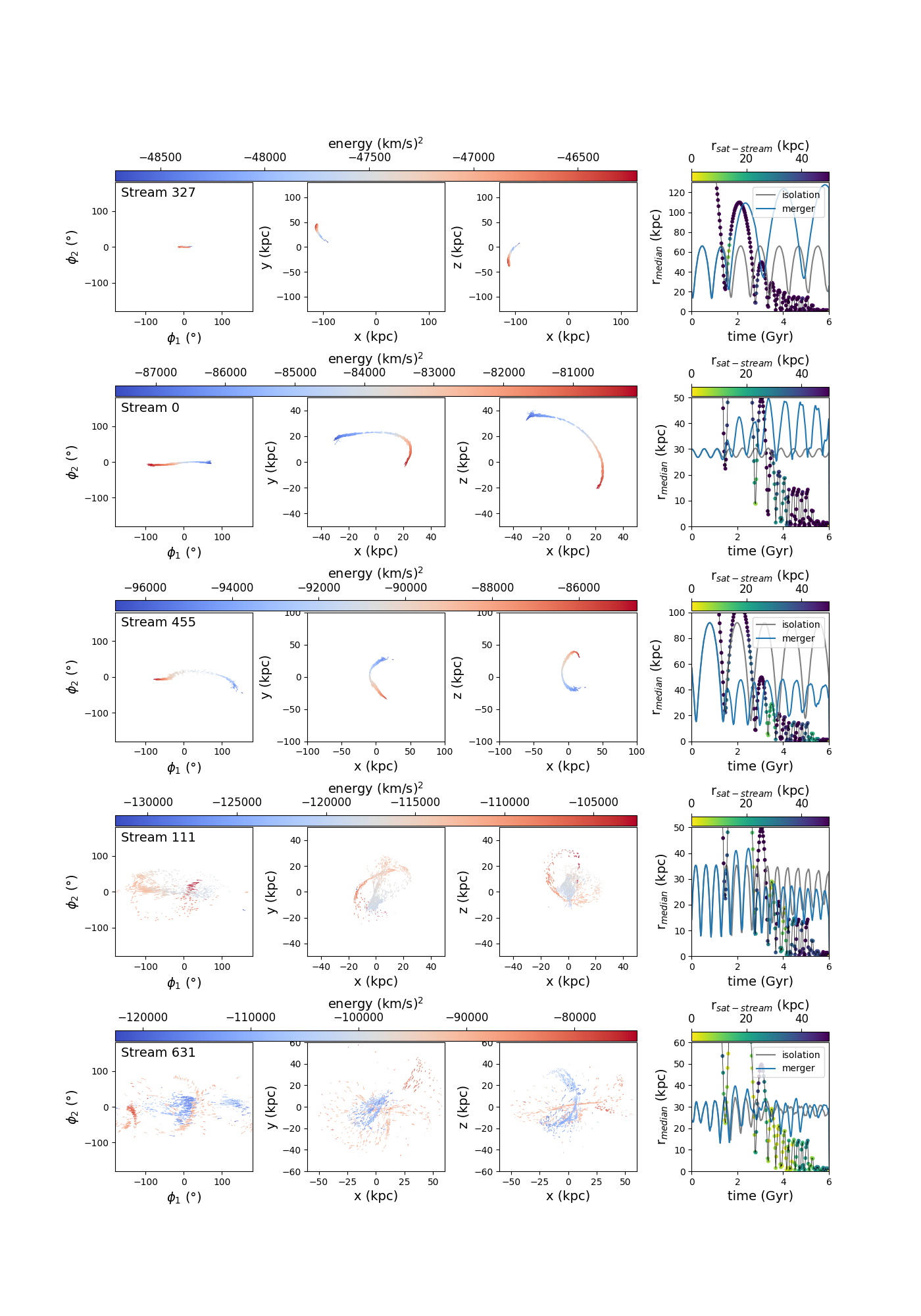}
    \caption{Circular orbit: Positions and velocities for star particles in 5 example stellar streams in their great circle coordinate frame (left column) and cartesian coordinates (x-y and x-z, middle columns colored by total energy, and their median radius with time (right column) during the merger (blue) and in the isolated halo (grey) and the perturber radius (black) with the distance between the perturber and the stream added (colored points).}
    \label{fig:r_time_circ}
\end{figure*}

\begin{figure*}[t]
    \centering
        \includegraphics[clip=true,trim={0cm 4cm 0cm 5cm}, width=0.95\textwidth]{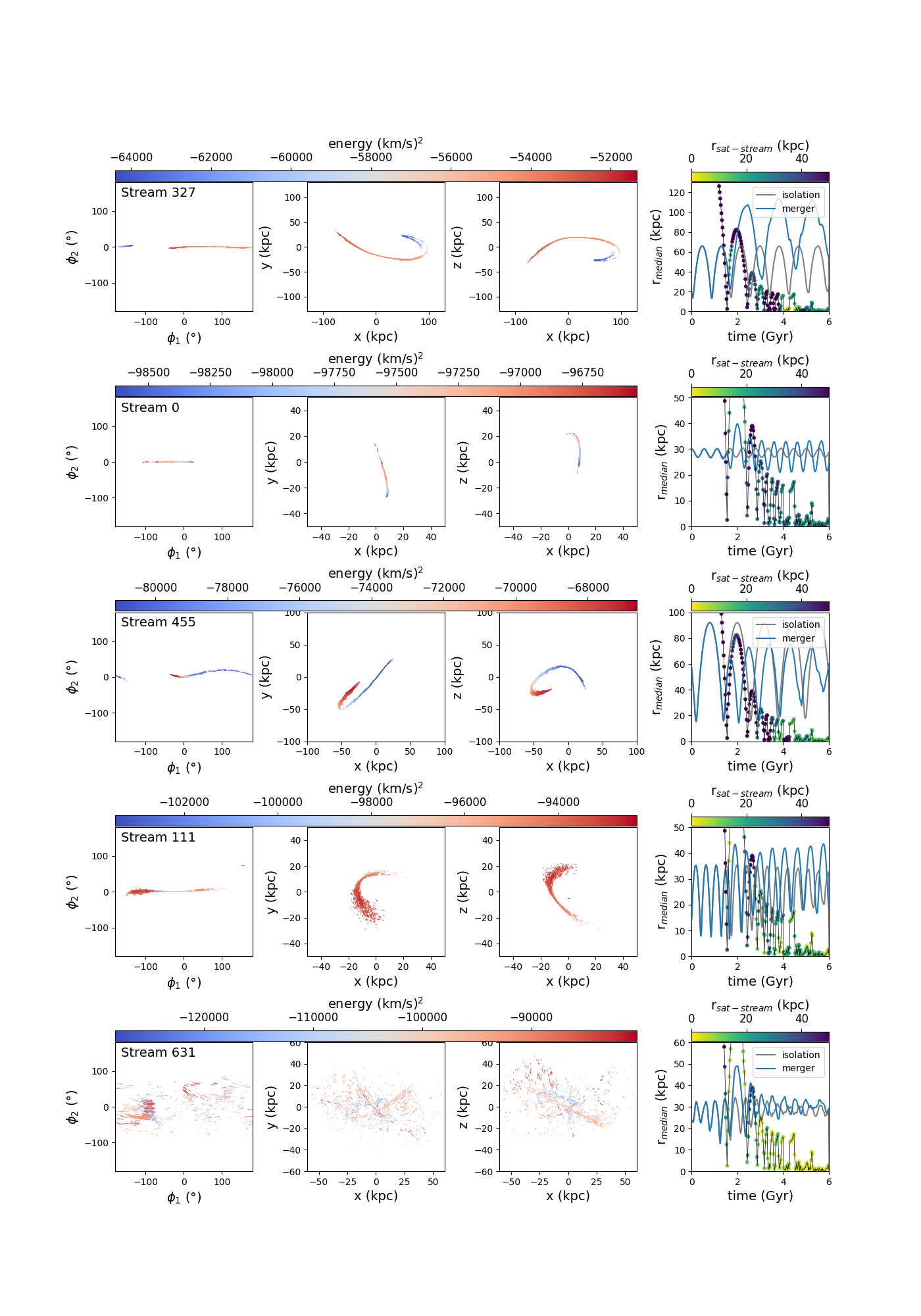}

    \caption{Radial orbit: Positions and velocities for star particles in 5 example stellar streams in their great circle coordinate frame (left column) and cartesian coordinates (x-y and x-z, middle columns colored by total energy, and their median radius with time (right column) during the merger (blue) and in the isolated halo (grey) and the perturber radius (black) with the distance between the perturber and the stream added (colored points).
    }
    \label{fig:r_time_radial}
\end{figure*}

In addition to the similarities in the changes in the energy distributions of the stellar stream, we also see statistically similar evolution of stream width and orbital plane as shown in Figures~\ref{fig:fig5}, \ref{fig:closest}, and \ref{fig:DeltaEL}. However, when comparing individual streams between different perturber orbits there are large differences in stream evolution. The exact same stellar streams are shown for the fiducial perturber orbit in Figure~\ref{fig:r_timecase}, for the more circular perturber orbit in Figure~\ref{fig:r_time_circ}, and for the radial perturber orbit in Figure~\ref{fig:r_time_radial}. While some streams see comparable behaviour (e.g. Stream 631 in the fifth row in each of Figures~\ref{fig:r_timecase}, \ref{fig:r_time_circ}, and \ref{fig:r_time_radial}), others exhibit opposite effect, with for example Stream 0 in the second row moving to larger radii in Figure~\ref{fig:r_time_circ} but not in Figures~\ref{fig:r_timecase} or Figure~\ref{fig:r_time_radial}, and Stream 455 in the third row moving to smaller radii in Figure~\ref{fig:r_timecase} and Figures~\ref{fig:r_time_circ} but not in Figure~\ref{fig:r_time_radial}.

\end{document}